\documentclass[aps, twocolumn, showpacs, preprintnumbers, amsmath, amssymb, prb]{revtex4-2}
\usepackage{multirow} % for cmd 'multirow', 'multicolumn'
\usepackage{tabularx}
\newcolumntype{Y}{>{\centering\arraybackslash}X}
\usepackage{tikz}
\usepackage{amsfonts}
\usepackage{amstext}
\usepackage{amsmath}
\usepackage{amssymb}
\usepackage{comment}
\usepackage{latexsym}
\usepackage{graphicx,epstopdf}
\usepackage{ulem}
\usepackage[colorlinks,linkcolor=blue,anchorcolor=blue,citecolor=blue]%
{hyperref}
\usepackage{times}
\usepackage{subfigure}
\usepackage{color}
\usepackage{dcolumn}
\usepackage{graphicx}
\usepackage{soul}
\usepackage{bbding}
\providecommand{\U}[1]{\protect\rule{.1in}{.1in}}
\newcommand{\CV}[0]{\color{black}}
\newcommand{\CIV}[0]{\color{black}}

\begin{document}
	\title{The second-order intrinsic Wiedemann-Franz law}
	\author{Ying-Fei Zhang$^{1}$}
	\author{Zhi-Fan Zhang$^{2}$}
	\author{Zhen-Gang Zhu$^{3,1}$}
	\email{zgzhu@ucas.ac.cn}
	\author{Gang Su$^{4,5}$}
	\email{gsu@ucas.ac.cn}
	\affiliation{
		$^{1}$ School of Physical Sciences, University of Chinese Academy of Sciences, Beijing 100049, China. \\
		$^{2}$ Interdisciplinary Center for Theoretical Physics and Information Sciences (ICTPIS), Fudan University, Shanghai 200433, China.\\
		$^{3}$ School of Electronic, Electrical and Communication Engineering, University of Chinese Academy of Sciences, Beijing 100049, China.\\
		%$^{3}$ CAS Center for Excellence in Topological Quantum Computation, University of Chinese Academy of Sciences, Beijing 100049, China.\\
		$^{4}$ Institute of Theoretical Physics, Chinese Academy of Sciences, Beijing 100190, China.\\
		$^{5}$ Kavli Institute for Theoretical Sciences, University of Chinese Academy of Sciences, Beijing 100190, China.
	}
	
	%	\pacs{}
	
	\begin{abstract}
		In recent years, the nonlinear anomalous thermal Hall effect has attracted substantial attention. In this paper, we carry out a theoretical exploration of the intrinsic anomalous thermal Hall and Nernst effect that is induced by the thermal Berry connection polarizability. This effect is independent of the relaxation time and can be present in antiferromagnets possessing $\mathcal{PT}$ symmetry. Additionally, we put forward a second-order intrinsic Wiedemann-Franz law, which represents the ratio of the second-order intrinsic thermal conductivity \CV coefficient \CIV to the second-order intrinsic electrical conductivity \CV coefficient \CIV. When analyzed within a four-band $\mathcal{PT}$ symmetric Dirac model, we observe that the second-order intrinsic thermal conductivity \CV coefficient \CIV is \CV linearly proportional to \CIV the second-order intrinsic electrical conductivity \CV coefficient \CIV, and the second-order intrinsic Wiedemann-Franz law is characterized by the chemical potential $\mu$ in the low-temperature regime. These findings provide significant implications for experimental verification.
	\end{abstract}
	%\revised{\today}
	
	\startpage{1}
	\endpage{ }
	\maketitle
	
	\section{Introduction}
	
	%\textit{Introduction.-} %The Hall effect \cite{Hall1879} represents a well-established phenomenon in which the imposition of an external magnetic field upon a conductor gives rise to a transverse current.
	%
	% Notably, the anomalous Hall effect (AHE) \cite{Nagaosa2010, Shen2017, Xiao2010} entails the generation of a transverse current that transpires without the need for any external magnetic field.
	%
	The Berry curvature (BC) $\boldsymbol{\Omega}$ \cite{Shen2017, Xiao2010}, likened to a magnetic field within the parameter space, % furnishing a potent framework for expounding the global characteristics of a system,
	plays a central role in modern quantum physics. %It can be , with Berry connection serving as an analogue to the vector potential.
	%
	%
	%When exposed to an electric field {$\bold E$}, an electron can attain an anomalous velocity \cite{Karplus1954, Kohn1957,Adams1959} that is directly proportional to $\bold E \times \boldsymbol\Omega$, 	thereby inducing an intrinsic Hall current \cite{Xiao2010}. The elaborate and complex interaction between the nontrivial topology of energy bands and the transverse Hall-like transport phenomena has been further clarified in the context of nonlinear response in the presence of anomalous velocity in recent years \cite{Battilomo2019,Low2015,You2018,Sodemann2015}.
	%
	It is found that the detailed distribution of BC in momentum space has profound impact on realistic physical world especially leading to topological nonlinearity.
	There exist the ohmic and Hall-type nonlinear currents  \cite{Tsirkin2022}. The ohmic current can be segmented into the extrinsic Drude current and the intrinsic current contributed by Berry connection polarizability (BCP) \cite{Wang2024,Kaplan2024,Jia2024,Nag2023}.
	The Hall type currents are instigated by Berry curvature dipole (BCD) and BCP.
	Sodemann and Fu \cite{Sodemann2015} postulated that in systems with time reversal symmetry ($\mathcal{T}$) and broken inversion symmetry ($\mathcal{P}$), the second-order Hall conductivity is proportional to BCD and exhibits a linear dependence on the relaxation time $\tau$, thereby signifying its extrinsic nature.
	A diverse array of materials, such as bilayer or few-layer WTe$_2$ \cite{Du2018,Ma2018}, the ferroelectric-like metal LiOsO$_{3}$ \cite{Xiao2020}, Weyl semimetals \cite{Zeng2021,Zhang2018,Wang2022a}, and others \cite{Zhang2022,Akatsuka2024,Pang2024}, have been recognized as manifesting this phenomenon.
	On the other hand, the gauge-invariant correction to the Berry connection \cite{Gao2014,Gao2015,Liu2021,Wang2021} also gives rise to a nonlinear anomalous Hall effect (NAHE) that is independent of $\tau$.
	This intrinsic effect emanates from BCP \cite{Liu2021,Wang2021,Liu2022,Pal2024} and preponderates in parity-inversion time-reversal ($\mathcal{PT}$)-symmetric materials since BCD vanishes due to symmetry constraints \cite{Zhang2023}.
	It is sensitively contingent upon the N\'{e}el vector \cite{Liu2021,Wang2021}, presenting a means for the electrical measurement of antiferromagnetic  materials.
	When the driving electric field is supplanted by a thermal flux, the associated intrinsic thermal transport is consequently associated with thermal Berry connection polarizability (TBCP) \cite{Li2024}, which represents an intrinsic thermal transport induced by nontrivial topological properties.
	
	Inspired by NAHE, the nonlinear anomalous thermal Hall effect (NATHE) and the nonlinear anomalous Nernst effect (NANE) have also attracted significant attention. These effects inherently give rise to corresponding anomalous electric, thermal, and thermoelectric transport coefficients, which have been the subject of extensive research \cite{Zhou2022,Yu2021,Saha2017,Choudhari2019,RoyKarmakar2022,Nandy2019,McCormick2017,Bhalla2021,Wang2023}.
	The relationships among these coefficients are encapsulated within the Wiedemann-Franz (WF) law and the Mott relation \cite{Qiang2023,Zeng2020,Nakata2022,Wang2022}. According to linear response theory \cite{Mermin1976},
	the WF law displays the ratio of electrical conductivity $\sigma_{ab}$ to thermal conductivity $\kappa_{ab}$ at low temperature as $\kappa_{ab}/\sigma_{ab} = LT$, where $L=\frac{1}{3}\left(\frac{k_{B}\pi}{e}\right)^{2}$ is the Lorentz number.
	Meanwhile, the Mott relation stipulates  $\alpha_{ab} = eLT \frac{\partial \sigma_ {ab}}{\partial \mu}$ with Nernst coefficient $\alpha_{ab}$ and chemical potential $\mu$.
	%
	%When extending these principles to the nonlinear regime, the second-order expressions of both the
	Recently, it is proposed that there exist nonlinear WF law and the Mott relation induced by BCD \cite{Wang2022},  %have been derived. When BCD plays a crucial role,
	in which the second-order WF law is given $\sigma_{abc}$ $\propto$ $\frac{\partial\kappa_{abc}}{\partial\mu}$, and the second-order Mott relation reads $\sigma_{abc}$ $\propto$ $\alpha_{abc}$ at low temperature, where $\sigma_{abc}$, $\kappa_{abc}$ and $\alpha_{abc}$ stand for the second-order conductivity, thermal conductivity and Nernst coefficient,  respectively.
	However, when TBCP prevails in $\mathcal{PT}$-symmetric materials, the above BCD-induced second-order WF law and Mott relation will be vanishing, and
	%relationships among the intrinsic nonlinear transport coefficients present an interesting conundrum, which is beneficial for analyzing and predicting the
	the intrinsic topological nonlinear thermoelectric  responses will be leading order, indicating the elaborate and complex interaction between the nontrivial topology
	of energy bands and the transverse nonlinear Hall-like transport. % phenomena has been further clarified in the context of nonlinear response

	In this work, we deduce formulas for the second-order intrinsic thermal and charge currents in the presence of a temperature gradient. These formulas are independent of the relaxation time  and are preponderant in $\mathcal{PT}$-symmetric materials.
	Subsequently, we present the second-order intrinsic WF law, which represents a ratio between the second-order intrinsic thermal conductivity \CV coefficient \CIV and the second-order intrinsic electrical conductivity \CV coefficient\CIV.
	To enhance our understanding of the impact of TBCP on nonlinear transport, we investigate a four-band Dirac model with $\mathcal{PT}$ symmetry. It is observed that when this intrinsic effect is dominant, the second-order intrinsic WF law $\kappa_{abc} \CV \sim \CIV \mu \sigma_{abc}$,  which is determined by $\mu$.

	\section{Theoretical formulation}
	%\textit{Theoretical formulation.-}
	The Hamiltonian including the effect of temperature gradient $\boldsymbol\nabla T$ is $\hat{H}_{T}=\hat{H}_0+\hat{H}'$ \cite{Smrcka1977}, where $\hat{H}_0$ is the Hamiltonian without perturbation, and
	\begin{equation}
		\begin{aligned}
			\hat{H}'=-\frac12\left[\hat{H}_0,\hat{\mathbf{r}}\right]_{+}\cdot \mathbf{E}_T,
		\end{aligned}
	\end{equation}
	where the subscript $+$ means anti-commutation, $\hat{\mathbf{r}}$ is position operator, and the thermal field $\mathbf{E}_T$ represents temperature gradient $\mathbf{E}_T=-\frac{\boldsymbol{\nabla} T}T$, where $\boldsymbol\nabla=\partial/\partial \boldsymbol{r}$.
	Using time-independent perturbation theory, the first-order $\mathcal{O}(\mathbf{E}_T)$ perturbed Bloch state \cite{Li2024} reads
	\begin{equation}
		\begin{aligned}
			|\tilde{u}_n(\boldsymbol{k})\rangle	=|u_n(\boldsymbol{k})\rangle-\sum_{l\neq n}\frac{\mathbf{E}_T\cdot\boldsymbol{\mathcal{A}}_{ln}(\boldsymbol{k})}{2}\frac{\varepsilon_{n,\boldsymbol{k}}+\varepsilon_{l,\boldsymbol{k}}}
			{\varepsilon_{n,\boldsymbol{k}}-\varepsilon_{l,\boldsymbol{k}}}|u_l(\boldsymbol{k})\rangle,
		\end{aligned}
	\end{equation}
	where $|u_n(\boldsymbol{k})\rangle$ is the unperturbed Bloch state, $\boldsymbol{\mathcal{A}}_{ln} (\boldsymbol{k})=\left\langle u_l(\boldsymbol{k})\right|i\boldsymbol{\nabla}_{\boldsymbol{k}}\left | u_n(\boldsymbol{k})\right\rangle$ is the interband Berry connection, where $\boldsymbol{\nabla}_{\boldsymbol{k}}=\partial/\partial \boldsymbol{k}$. $\varepsilon_{n, \boldsymbol{k}}$ is the unperturbed band energy and $n$, $l$ are band indexes.

	The perturbed band energy $\tilde{\varepsilon}_{n, \boldsymbol{k}}$ is %to the second order in $\mathbf{E}_T$ is
	\begin{equation}
		\begin{aligned}
			\label{eq:energy_correction}
			\tilde{\varepsilon}_{n,\boldsymbol{k}}&=\varepsilon_{n,\boldsymbol{k}}+\varepsilon_{n,\boldsymbol{k}}^{(1)}+\varepsilon_{n,\boldsymbol{k}}^{(2)},
		\end{aligned}
	\end{equation}
	where the first-order $\mathcal{O}(\mathbf{E}_T)$ correction to the band energy vanishes \cite{Gao2014,Nag2023}, i.e., $\varepsilon_{n,\boldsymbol{k}}^{(1)}=\langle u_{n}(\boldsymbol{k})|\hat H^{\prime}|u_{n}(\boldsymbol{k})\rangle=0$, and $\varepsilon_{n,\boldsymbol{k}}^{(2)}$ is proportional to $\mathbf{E}_T^2$.
	The perturbed BC $\tilde{\boldsymbol{\Omega}}_n({\boldsymbol{k}})$ reads
	\begin{equation}
		\label{eq:BC_correction}
		\tilde{\boldsymbol{\Omega}}_n({\boldsymbol{k}})=\boldsymbol{\Omega}_n(\boldsymbol{k})+\boldsymbol{\Omega}^{(1)}_n(\boldsymbol{k}), %\quad
		\boldsymbol{\Omega}^{(1)}_n(\boldsymbol{k})=\nabla_{\boldsymbol{k}}\times\boldsymbol{\mathcal{A}}_n^{(1)}(\boldsymbol{k}),
	\end{equation}
	where $\boldsymbol{\Omega}_n(\boldsymbol{k})$ is the unperturbed BC, and $\boldsymbol{\Omega}^{(1)}_n(\boldsymbol{k})$ is the first-order $\mathcal{O}(\mathbf{E}_T)$ correction of BC, $\boldsymbol{\mathcal{A}}_n^{(1)}(\boldsymbol{k})$ is the first-order Berry connection, which effectively captures the band geometric quantity, and takes the form \cite{Li2024}
	\begin{equation}
		\begin{aligned}
			\mathcal{A}_{n,a}^{(1)}(\boldsymbol{k})=\mathbb{G}_{n,ab}^{\text{T}}(\boldsymbol{k})\mathbf{E}_{T,b}.\\
		\end{aligned}
	\end{equation}
	The Einstein summation for $a$ and $b$ is implicit, and indices $a$ and $b$ denote the Cartesian coordinates, $\mathbf{E}_{T,b}$ is the component of the thermal field $\mathbf{E}_T$ in the direction $b$,
	$\mathbb{G}_{n}^\text{T}(\boldsymbol{k})$ is the TBCP tensor and its components are defined as
	\begin{equation}
		\begin{aligned}
			\label{eq:G_T}
			\mathbb{G}_{n,ab}^{\text{T}}(\boldsymbol{k})=-\Re %\left\{
			\sum_{l\neq n}
			\frac{\left(\varepsilon_{n,\boldsymbol{k}}+\varepsilon_{l,\boldsymbol{k}}\right){\mathcal{A}}_{nl,a}(\boldsymbol{k}){\mathcal{A}}_{ln,b}(\boldsymbol{k})}
			{\varepsilon_{n,\boldsymbol{k}}-\varepsilon_{l,\boldsymbol{k}}}. %\right\}.
		\end{aligned}
	\end{equation}
	
	With regard to the NAHE, the electric current expands into linear and nonlinear terms according to the electric field, $j_a^E=\sigma_{ab}E_b+\sigma_{abc}E_bE_c+\cdots$.
	Considering the NATHE, the thermal current is expanded in powers of temperature gradient, $j_a^Q=\kappa_{ab}\CV(-\nabla_bT)\CIV+\kappa_{abc}\CV(-\nabla_bT)(-\nabla_cT)\CIV+\cdots$. For the NANE, the electric current is expressed as a series expansion in terms of the temperature gradient, i.e. $j_a^N= \alpha_{ab}\CV(-\nabla_bT)\CIV+ \alpha_{abc}\CV(-\nabla_bT)(-\nabla_cT)\CIV+\cdots$, where $\{a,b,c\}\in\{x,y,z\}$ represents the Cartesian coordinates.
	%
	%There are coefficients between the response currents and the applied fields, and we can define the linear (second-order) electrical conductivity $\tilde\sigma_{ab}$ ($\tilde\sigma_{abc}$), the linear (second-order) thermal conductivity $\tilde\kappa_{ab}$ ($\tilde\kappa_{abc}$), and the linear (second-order) Nernst coefficient $\tilde L_{ab}$ ($\tilde L_{abc}$).
	%
	%The reason that we expand the transport coefficients to higher orders is because of the symmetry constraints in some cases, the linear response disappears while the nonlinear response predominates.
	%
	Symmetry constraints may prohibit linear response so that the nonlinear response predominates.
	Such as, when time-reversal symmetry $\mathcal{T}$ is broken, the linear response dominates due to Onsager's reciprocity relations \cite{ Du2021}; when $\mathcal{T}$ is preserved but inversion symmetry $\mathcal{P}$ is broken, the linear response vanishes, and the second-order response of the BCD mechanism becomes dominant \cite{Sodemann2015}.
	However, when $\mathcal{T}$ and $\mathcal{P}$ are individually broken, while $\mathcal{PT}$ symmetry is preserved, both the linear response and the BCD mechanism disappear,
	and the intrinsic second-order response of the (T)BCP mechanism becomes dominant \cite{Liu2021,Wang2021,Li2024}, which is the main focus of our work.

	\CV\subsection{Second-order intrinsic anomalous thermal Hall effect}\CIV\label{IATHE}
	%\textit{Intrinsic second-order anomalous thermal Hall effect.-}
	The total anomalous thermal current $\boldsymbol{j}^{{Q}}_{\text{total}}$ can be decomposed into three components: $ \boldsymbol{j}^Q_{v}$, $\boldsymbol{j}^Q_{E}$, and $\boldsymbol{j}^Q_{T}$.
	%
	%Each component
	%representing distinct physical mechanisms. %contributing to the thermal current.
	%
	Specifically, $\boldsymbol{ j}^Q_{v}$ refers to the  longitudinal thermal current caused by group velocity $\boldsymbol{v}_{\boldsymbol{k}}$ of carriers,
	while $\boldsymbol{j}^Q_{E}$ is transverse thermal current driven by an external electric field $\mathbf{E}$ and is influenced by the nontrivial Berry curvature $\boldsymbol{\Omega}_n(\boldsymbol{k})$.
	In this paper, we focus on $\boldsymbol{j}^Q_{T}$, which is the transverse thermal current mediated by the nontrivial Berry curvature in the presence of $\boldsymbol\nabla T$ with zero electromagnetic field, given by  \cite{Yokoyama2011,Bergman2010}
	\begin{eqnarray}
		%\begin{aligned}
		%
		\boldsymbol{j}^{Q}_{T} &=& -\frac{k_B^2T}\hbar\boldsymbol\nabla T\times\int[d\boldsymbol{k}]\sum_n\boldsymbol{\Omega}_n(\boldsymbol{k})\bigg[\beta^2\big(\varepsilon_{n,\boldsymbol{k}}-\mu\big)^2f_0^n \notag\\
		&+& \frac{\pi^2}3-\ln^2(1-f_0^n)-2\operatorname{Li}_2(1-f_0^n)\bigg],
		%\end{aligned}
		\label{eq:j_Q_T}
	\end{eqnarray}
	where \CV$\times$ stands for cross product operation\CIV, $[d\boldsymbol{k}]$ stands for $d^D k/(2\pi)^D$, $D$ is the dimension, the Fermi-Dirac distribution function is expressed as ${f}_{0}^n=1/[1+e^{\beta({\varepsilon_{n,\boldsymbol{k}}}-\mu)}]$, with $\beta=\frac{1}{k_BT}$, and $\text{Li}_n(x)=\sum_{m=1}^\infty\frac{x^m}{m^n}$ is the polylogarithm function.

	When considering the influence of $\boldsymbol\nabla T$ on transport properties with $\mathcal{PT}$ symmetry, the thermal current can be derived by substituting the expressions $\tilde{\varepsilon}_{n,\boldsymbol{k}}$ and $\tilde{\boldsymbol{\Omega}}_n(\boldsymbol{k})$ from Eqs. (\ref{eq:energy_correction}) and (\ref{eq:BC_correction}) into Eq. (\ref{eq:j_Q_T}).
	The second-order current is obtained by selectively collecting the terms proportional to $(\nabla T)^2$, and it is evident that only the first-order correction of Berry curvature $\boldsymbol{\Omega}^{(1)}_n(\boldsymbol{k})$ contributes to the second-order intrinsic transverse thermal current, implying that the thermal current arises as a result of modulated TBCP.
	%
	%The second-order intrinsic anomalous thermal conductivity
	\CV The response tensor \CIV $\kappa_{abc}$  is defined by
	\begin{equation}
		\begin{aligned}
			j_{T,a}^{Q,(2)}=\kappa_{abc}\CV(-{\nabla_b T})(-{\nabla_c T})\CIV,
		\end{aligned}
	\end{equation}
	where \CV  the superscript (2) represents the second-order term,  so does the following, and
	\CIV
	\begin{align}
		\label{eq:kappa_abc_unsimply}
		\kappa_{abc}=&\frac{k_{B}^{2}}{\hbar}\int[d\boldsymbol{k}]\sum_n\Lambda^n_{abc}(\boldsymbol{k}),
	\end{align}
	with integrand \CV (see Appendix \ref{app:k_abc_unsimply}) \CIV
	\begin{align}
		\label{eq:integrand}
		\Lambda_{abc}^n(\boldsymbol{k})
		& = \epsilon_{ab}\partial_{a}\mathbb{G}^{\text{T}}_{n,bc}(\boldsymbol{k})\bigg[\frac{{(\varepsilon_{{n,\boldsymbol{k}}}-\mu)^{2}}f_{0}^n}{(k_{B}T)^{2}}\nonumber\\
		& +\pi^2/3-\ln^2(1-f_0^n)-2\mathrm{Li}_2(1-f_0^n)\bigg],
	\end{align}
	where $\epsilon_{\alpha\beta}$ is Levi-Civita antisymmetric tensor, $\partial_{a}={\partial}/{\partial k_a}$.
	It is seen that this second-order thermal Hall effect is independent of $\tau$ and only related to band structure.

	\CV\subsection{Second-order intrinsic anomalous Nernst effect}	\CIV
	%\textit{Intrinsic second-order amomalous Nernst effect.-}
	The NANE refers to the generated transverse charge current by applied $\boldsymbol\nabla T$ due to Berry curvature  \cite{Xiao2006}
	\begin{equation}
		\boldsymbol{j}^{N}   = -\frac{e}{\hbar}\boldsymbol{\nabla}T\times\sum_m\int[d{\boldsymbol{k}}]\boldsymbol{\Omega}_m(\boldsymbol{k})  S_{m}(\mathbf{r},\boldsymbol{k}), %\frac{(\varepsilon_{m,\boldsymbol{k}}-\mu)}Tf_0^m \right.\nonumber\\
		%	&\left. +k_B\ln\left(1+e^{-\beta(\varepsilon_{m,\boldsymbol{k}}-\mu)}\right) \right ].
		\label{eq:J_N_E}
	\end{equation}
	$S_{m}(\mathbf{r},\boldsymbol{k})=\frac{(\varepsilon_{m,\boldsymbol{k}}-\mu)}Tf_0^m+k_B\ln\left(1+e^{-\beta(\varepsilon_{m,\boldsymbol{k}}-\mu)}\right) $ is entropy density.
	Analogous to the derivation of $\kappa_{abc}$, we can derive the  Nernst-like current by substituting the expressions $\tilde{\varepsilon}_{n,\boldsymbol{k}}$ and $\tilde{\boldsymbol{\Omega}}_n(\boldsymbol{k})$ from Eqs. (\ref{eq:energy_correction}) and (\ref{eq:BC_correction}) into Eq. (\ref{eq:J_N_E}), suggesting that the intrinsic Nernst current is generated due to TBCP.
	The second-order intrinsic Nernst-like current is denoted as
	\begin{equation}
		j_{a}^{N,\mathrm{(2)}} = \alpha_{abc}\CV(-{\nabla_b T})(-{\nabla_c T})\CIV,
		\label{jEalphaTBCP2}
	\end{equation}
	and the corresponding second-order Nernst coefficient $\alpha_{abc}$ can be simplified as %to Eq. (\ref{eq:L_abc})
	\CV(see Appendix \ref{app:Derivation of L})\CIV,
	\begin{equation}
		\alpha_{abc} = \frac{e}{\hbar T}\epsilon_{ab}\sum_m\int[d{\boldsymbol{k}}]
		\partial_{a}\mathbb{G}^{\text{T}}_{m,bc}(\boldsymbol{k}) S_{m}(\mathbf{r},\boldsymbol{k}).
		%\bigg[(\varepsilon_{{m,\boldsymbol{k}}}-\mu)f_{0}^m  \notag\\
		%&+& k_BT\ln(1+e^{-\beta(\varepsilon_{m,\boldsymbol{k}}-\mu)})\bigg],
		\label{eq:L_abc}
	\end{equation}
	%{\color{blue}through the application of integration by parts, it becomes evident that the second order Nernst coefficient depends on the $f_0$ term, so it is related to ``Fermi sea'' effect}.% meaning that the Nernst coefficient still exists in the band gap of a material.}

\CV\subsection{Second-order intrinsic  Wiedemann-Franz law}\CIV
%\textit{Intrinsic second-order Wiedemann-Franz law.- }
By solving the Boltzmann equation, it is found that the second-order conductivity consists of three components \cite{Zhang2023,Kaplan2023}, $\sigma_{abc}=\sigma^{\text{Drude}}_{abc}+\sigma^\text{BCP}_{abc}+\sigma^\text{BCD}_{abc}$, in which \CV$\sigma^{\text{Drude}}_{abc}=-\frac{e^3\tau^2}{\hbar^3}\int[dk]\partial_a \varepsilon_{n,\boldsymbol k}\partial_b\partial_cf_0^n$ \cite{Tsirkin2022} \CIV and \CV$\sigma^\text{BCD}_{abc}=-\frac{\tau e^3}{\hbar^2}\epsilon_{abd}\int[dk]\Omega_{n, d}(k)\partial_cf_0^n$ \cite{Sodemann2015} \CIV are both extrinsic contributions, related to $\tau^2$ and $\tau$ respectively.
In this study, we focus on the second intrinsic term \cite{Liu2021,Wang2021}
\begin{align}
	\label{conductivity-BCP}
	\sigma^\text{BCP}_{abc}=&-\frac{e^2}{\hbar}\epsilon_{ab}\sum_m\int[d{\boldsymbol{k}}]\partial _a \mathbb{G}^{\text{E}}_{m,bc}(\boldsymbol{k})f_{0}^m,
\end{align}
where $\mathbb{G}_{m,ab}^{\text{E}}(\boldsymbol{k})$ is the BCP tensor \cite{Gao2014,Lai2021},
\begin{equation}
	%\begin{aligned}
	\mathbb{G}_{m,ab}^{\text{E}}(\boldsymbol{k})=2e\Re\sum_{m^{\prime}\neq m}\frac{\mathcal{A}_{mm^{\prime},a}(\boldsymbol{k})\mathcal{A}_{m^{\prime}m,b}(\boldsymbol{k})}{\varepsilon_{m,\boldsymbol{k}}-\varepsilon_{m^{\prime},\boldsymbol{k}}}.
	%\end{aligned}
	\label{GmabE}
\end{equation}
Eq. (\ref{conductivity-BCP}) shows the ``Fermi surface'' effect for NAHE, where only the bands near Fermi energy contribute to the integration.

%Further, we obtain the second-order intrinsic Wiedemann-Franz law, describing the relationship between the second-order intrinsic thermal conductivity and second-order intrinsic electrical conductivity.
At low temperature \cite{RoyKarmakar2022,Yokoyama2011}, utilizing the Sommerfeld expansion \cite{Mermin1976}, \CV the second-order intrinsic WF law reads (see Appendix \ref{sec:A})\CIV
\begin{equation}
	\label{eq:wflaw}
	\kappa_{abc}=-\frac{\pi^2k_B^3T}{3e^3}\frac{\epsilon_{ab} P_{a,bc}^\text{T}(\mu)}{\epsilon_{ab}P_{a,bc}^\text{E}(\mu)}\sigma_{abc},
\end{equation}
where  $P^{\text{T}}_{a,bc}$ and $P^{\text{E}}_{a,bc}$ are defined as
\begin{equation}
	%\begin{aligned}
	P^{\text{T(E)}}_{a,bc}(\mu) =  \frac{1}{k_BT(e)}\sum_n\int[d{\boldsymbol{k}}]\mathbb{G}^{\text{T(E)}}_{n, bc}(\boldsymbol{k})\delta(\mu-\varepsilon_{n,{\boldsymbol{k}}})\frac{\partial\varepsilon_{n, {\boldsymbol{k}}}}{\partial k_{a}}. %\notag\\
	%	
	%  P^{\text{E}}_{a,bc}(\varepsilon) &=& \frac{1}{e}\sum_n\int[d{\boldsymbol{k}}]G_{n,bc}^{{\text{E}}}\delta(\varepsilon-\varepsilon_{{n,\boldsymbol{k}}})\frac{\partial\varepsilon_{{n,\boldsymbol{k}}}}{\partial k_{a}}.
	%\end{aligned}
	\notag
	%\label{eq:P_T}
\end{equation}

%%%%%%%%%%%%%%%%%%%%%%%%%%%%%%%%%%%%%%%%%%%%%%%%%%%%%%%%%%%%%%%%%%%%%%%%%%%%%%%%%%%%%%%%%%%%%%%%%%%%%%%%%%
\begin{figure}[tb]
	\centering
	\includegraphics[width=1\linewidth]{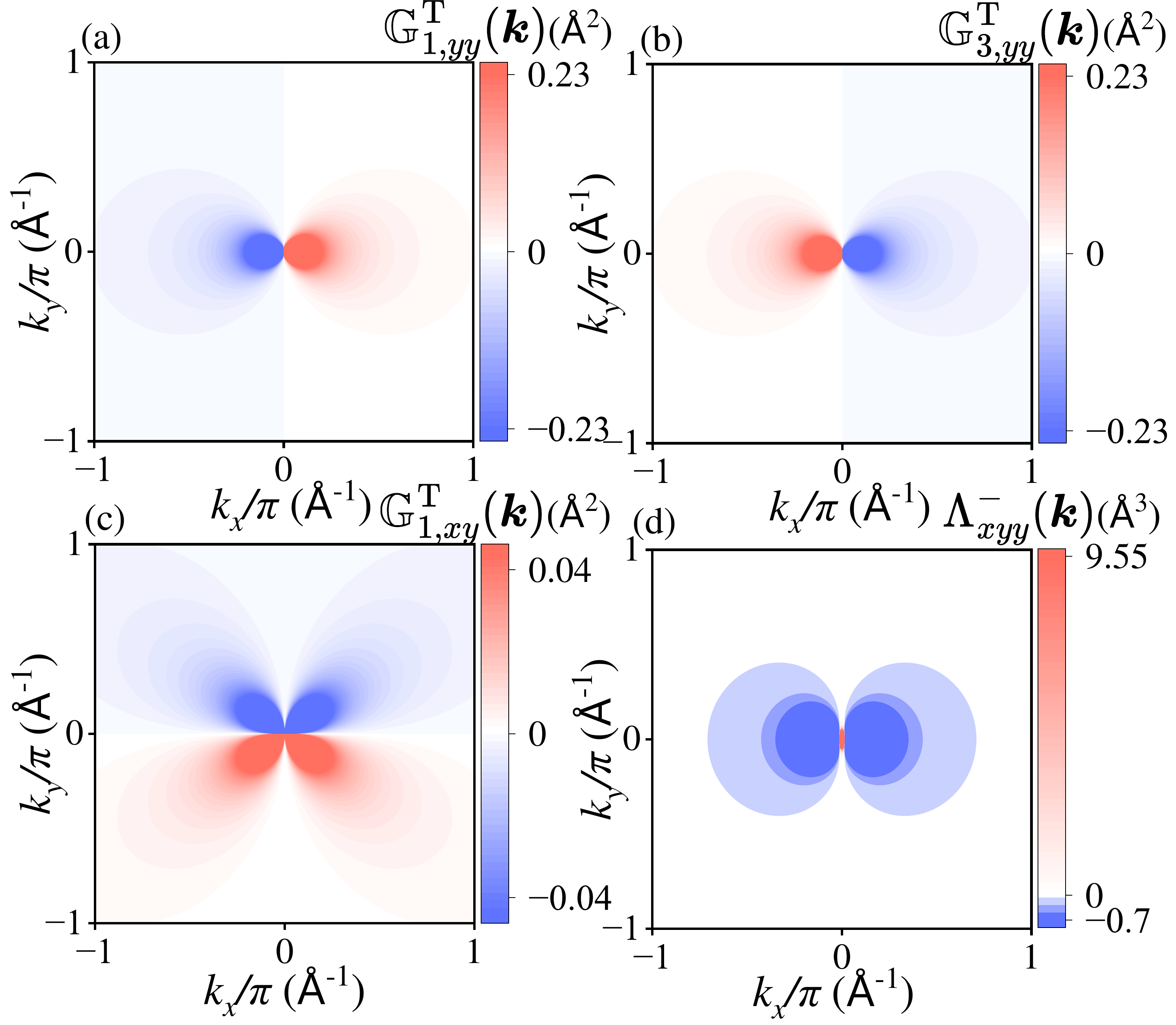}
	\caption{(a) The contour plots of TBCP \CV$\mathbb{G}^{\text{T}}_{1,yy}(\boldsymbol{k})$ \CIV in momentum space for the valence bands. (b) The contour plots of TBCP \CV$\mathbb{G}^{\text{T}}_{3,yy}(\boldsymbol{k})$\CIV in momentum space for the conduction bands. (c) The contour plots of TBCP \CV$\mathbb{G}^{\text{T}}_{1,xy}(\boldsymbol{k})$ \CIV in momentum space for the valence bands. (d) The integrated function of thermal conductivity \CV$\Lambda^{-}_{xyy}(\boldsymbol{k})$\CIV for the valence bands.  We set $v_x=v_y=1$ eV$\mathring{\text{A}}$, $\Delta=40$ meV, $k_B T=1$ meV ($T\sim$11.6 K), $w/v_x=0.4$. }
	\label{G_yy}
\end{figure}
%%%%%%%%%%%%%%%%%%%%%%%%%%%%%%%%%%%%%%%%%%%%%%%%%%%%%%%%%%%%%%%%%%%%%%%%%%%%%%%%%%%%%%%%%%%%%%%%%%%%%%%%%%

%%%%%%%%%%%%%%%%%%%%%%%%%%%%%%%%%%%%%%%%%%%%%%%%%%%%%%%%%%%%%%%%%%%%%%%%%%%%%%%%%%%%%%%%%%%%%%%%%%%%%%%%%%
%%%%%%%%%%%%%%%%%%%%%%%%%%%%%%%%%%%%%%%%%%%%%%%%%%%%%%%%%%%%%%%%%%%%%%%%%%%%%%%%%%%%%%%%%%%%%%%%%%%%%%%%%%
\section{A model study}
%\textit{A model study.-}

\begin{table*}[ht]\CV
	\centering
	\caption{The symmetry properties of the thermoelectric transport (a) coefficients of linear order, second-order extrinsic (induced by  BCD) and second-order intrinsic (induced by BCP/TBCP) under spatial inversion ($\mathcal{P}$), time reversal ($\mathcal{T}$ ), and the combined operation of spatial and time reversal ($\mathcal{PT}$) are presented. The allowed (forbidden) conductivities are indicated by \checkmark (\XSolidBrush ).}
	\begin{ruledtabular}
		\begin{tabular}{cccccc}
			Order&contributions& response coefficients & $\mathcal {P} $& $\mathcal {T} $& $\mathcal {PT}$ \\
			\hline
			\multirow{6}{*}{linear  }
			&\multirow{3}{*}{Drude ($\propto$ $\tau^1$) \cite{Mermin1976}}
			& $\sigma_{ab}^{\text{Drude}}$ &  & &  \\
			&& $\alpha_{ab}^{\text{Drude}}$ &\checkmark &  \checkmark& \checkmark\\
			&& $\kappa_{ab}^{\text{Drude}}$ & & & \\
			\\
			&\multirow{3}{*}{BC ($\propto$ $\tau^0$) \cite{Xiao2010,Shen2017}}
			& $\sigma_{ab}^{\text{BC}}$ &  & &  \\
			&& $\alpha_{ab}^{\text{BC}}$ &\checkmark &  \XSolidBrush & \XSolidBrush  \\
			&& $\kappa_{ab}^{\text{BC}}$ & & & \\
			\\\\
			\multirow{9}{*}{second} &\multirow{3}{*} {Drude ($\propto$ $\tau^2$) \cite{Tsirkin2022}} & $\sigma^{\text{Drude}}_{abc}$ &  & &  \\
			&& $\alpha^{\text{Drude}}_{abc}$ & \XSolidBrush  & \XSolidBrush &\checkmark \\
			&& $\kappa^{\text{Drude}}_{abc}$ & & & \\\\
			&\multirow{3}{*} {BCD ($\propto$ $\tau$) \cite{Sodemann2015, Zeng2020} } & $\sigma^{\text{BCD}}_{abc}$ &  & &  \\
			&& $\alpha^{\text{BCD}}_{abc}$ &  \XSolidBrush  & \checkmark &  \XSolidBrush \\
			&& $\kappa^{\text{BCD}}_{abc}$ & & & \\\\
			%	\cline{2-6}
			&\multirow{3}{*}{BCP  ($\propto$ $\tau^0$) \cite{Liu2021} / TBCP  ($\propto$ $\tau^0$) in our work
			}& $\sigma^{\text{BCP}}_{abc}$ & &  &  \\
			&& $\alpha_{abc}$ &   \XSolidBrush  &   \XSolidBrush &\checkmark \\
			&& $\kappa_{abc}$  & & & \\
		\end{tabular}
		\label{tab111}
	\end{ruledtabular}
\end{table*}
\CIV
\CV
Under time reversal operation, $\boldsymbol{\Omega}_{n}(\boldsymbol{k})$ is $\mathcal{T}$ odd, wave vector $\boldsymbol k$ is $\mathcal{T}$ odd; while under the spatial inversion operation, $\boldsymbol{\Omega}_{n}(\boldsymbol{k})$ is $\mathcal{P}$ even, the wave vector $\boldsymbol k$ is $\mathcal{P}$ odd. We can know that $\mathcal{PT}\boldsymbol{\Omega}_{n}(\boldsymbol{k})=-\boldsymbol{\Omega}_{n}(\boldsymbol{k})=\boldsymbol{\Omega}_{n}(\boldsymbol{k})$ in a $\mathcal{PT}$-symmetric system, which indicates the Berry curvature must be zero.
\CIV
Therefore, the Berry curvature $\boldsymbol{\Omega}_{n}(\boldsymbol{k})$ vanishes in materials that preserve $\mathcal{PT}$ symmetry, implying that linear and second-order responses contributed by BC are forbidden.
However, \CV both TBCP $\mathbb{G}_{n,ab}^{\text{T}}(\boldsymbol{k})$ and BCP $\mathbb{G}_{n,ab}^{\text{E}}(\boldsymbol{k})$ are $\mathcal{T}$ even under time reversal operation and $\mathcal{P}$ even under spatial inversion operation. That is to say $\mathcal{PT}\mathbb{G}_{n,ab}^{\text{T(E)}}(\boldsymbol{k})=\mathbb{G}_{n,ab}^{\text{T(E)}}(\boldsymbol{k})$, \CIV indicating the TBCP and BCP remain nonzero under the symmetry of $\mathcal{PT}$.  \CV To more effectively illustrate the symmetry of transport coefficients across different mechanisms, Table \ref{tab111} summarizes the symmetries of linear and second-order transport coefficients induced by various contributions: the Drude, the BC, the BCD terms ($\tau$-dependent), and BCP/TBCP terms ($\tau$-independent). From the Table \ref{tab111}, disregarding the Drude contribution (longitudinal, $\tau$ dependent), it is evident that the second-order intrinsic response plays a dominant role in systems with $\mathcal{PT}$ symmetry.
\CIV

To better understand and concretize the effect of intrinsic mechanism on nonlinear transport (no extrinsic mechanisms, only exists our proposed intrinsic mechanism), we consider a four-band Dirac model with $\mathcal{PT}$ symmetry \cite{Liu2021,Tang2016} %, indicating that every band is doubly degenerate \cite{Tang2016}
\begin{equation}
	\label{eq:model}
	\mathcal{H}=wk_x+v_xk_x\tau_x+v_yk_y\tau_y\sigma_x+\Delta\tau_z,
\end{equation}
where  \CV($k_x$, $k_y$) is the wave vector\CIV,  the $\tau$'s and $\sigma$'s are two sets of Pauli matrices, and $w$ and $2\Delta$ represent the tilt of energy band and the magnitude of gap, respectively. The parameters $v_x$, $v_y$ and $\Delta$ are real.
The energy spectrum is given by
\begin{equation}
	\begin{aligned}
		\label{eq:energy}	
		\varepsilon_{\pm, \boldsymbol{k}}=wk_x \pm \CV E_0(\boldsymbol k)\CIV,
	\end{aligned}
\end{equation}
where $\pm$ denotes the conduction and valence bands\CV, and $E_0(\boldsymbol k)=\sqrt{\Delta^2 + k_x^2 v^2 + k_y^2 v^2}$ represents the energy splitting between the two bands. The $\mathcal{PT}$ operation is expressed as  $-i\sigma_yK$, where $K$ corresponds to a
complex conjugation. By applying the $\mathcal{PT}$ operation, it can be confirmed that
$\mathcal{PT} \mathcal{H} (\mathcal{PT})^{-1}=\mathcal{H}$, thereby demonstrating the fulfillment of $\mathcal{PT}$ symmetry. Consequently, each energy band is required to be doubly degenerate \cite{Tang2016}.\CIV

In this model, $n=1, 2$ are for valence bands and $n=3, 4$ are for conduction bands. The analytical expression of the TBCP for the valence band is evaluated from Eq. (\ref{eq:G_T})
\begin{equation}
	\label{Gexpression}
	\mathbb{G}_1^{\text{T}}(\boldsymbol k)=\frac{wk_xv^{4}}{4\CV E_0^{5}(\boldsymbol k)\CIV}
	\begin{pmatrix}
		\frac{\Delta^{2}}{v^{2}}+k_y^{2}&-k_xk_y\\-k_xk_y&\frac{\Delta^{2}}{v^{2}}+k_x^{2}
	\end{pmatrix},
\end{equation}
\CV here\CIV, $v_x = v_y = v$ is assumed due to the isotropy in further analytical calculations.
The TBCP of conduction and valence bands are opposite in sign, $\mathbb{G}_1^{\text{T}}(\boldsymbol{k})=-\mathbb{G}_3^{\text{T}}(\boldsymbol{k})$.
The TBCP elements \CV$\mathbb{G}^{\text{T}}_{1,yy}(\boldsymbol{k})$ \CIV, \CV$\mathbb{G}^{\text{T}}_{3,yy}(\boldsymbol{k})$ \CIV  and \CV$\mathbb{G}^{\text{T}}_{1,xy}(\boldsymbol{k})$ \CIV  on the ($k_x$, $k_y$) plane are depicted in Fig. \ref{G_yy}(a)-(c), which are primarily concentrated  around the gap region.
The diagonal components of TBCP develop dipole-like structures \CV[\CIV Fig.\ref{G_yy}(a) and (b)\CV]\CIV; while the off-diagonal component exhibits a quadrupole-like structure in momentum space \CV[\CIV Fig. \ref{G_yy}(c)\CV ]\CIV.
And $\mathbb{G}^{\text{T}}_{1,yy}(\boldsymbol{k})=-\mathbb{G}^{\text{T}}_{3,yy}(\boldsymbol{k})$ is seen in Fig. \ref{G_yy}(a) and Fig. \ref{G_yy}(b).
\CV
Apparently, it is evident that both $\mathbb{G}^{\text{T}}_{1,yy}(\boldsymbol{k})$ and $\mathbb{G}^{\text{T}}_{3,yy}(\boldsymbol{k})$ are odd functions of $k_x$; while $\mathbb{G}^{\text{T}}_{1,xy}(\boldsymbol{k})$ in Fig. \ref{G_yy}(c) is odd along the $k_y$ axis, not $k_x$.
\CIV
Moreover, the integrand \CV$\Lambda^{-}_{xyy}(\boldsymbol{k})=\sum_{n=1,2}\Lambda_{xyy}^n(\boldsymbol{k})$  for second-order intrinsic thermal conductivity coefficient $\kappa_{xyy}$ on the valence bands \CIV is displayed in Fig. \ref{G_yy}(d), which is a combination of a monopole and a dipole along $k_x$.

The tilt term is important and induces non-zero TBCP and transport coefficients due to broken inversion symmetry. %, and an overall energy shift for both bands, as seen in Eqs. (\ref{eq:G_T}) and (\ref{Gexpression}).
According to Eqs. (\ref{eq:kappa_abc_unsimply}) and (\ref{eq:L_abc}), $\kappa_{abc}$ and $\alpha_{abc}$ are both antisymmetric with respect to first two indices, illustrating intrinsic Hall-type nature.
As a result, only two independent components need to be considered: $\kappa_{xyy}=-\kappa_{yxy}$ and $\kappa_{xyx}=-\kappa_{yxx}$, with all other terms vanishing, \CV and $\sigma_{abc}^{\text{BCP}}$ in Eq. (\ref{conductivity-BCP}) also satisfies the aforementioned conditions. \CIV

Taking $xyy$ component as an example, we obtain  the second-order intrinsic thermal conductivity \CV coefficient \CIV contributed by TBCP  \CV (see Appendix \ref{APP:WFLAW} for more details)\CIV
\CV
\begin{equation}	\label{kappajiexi}
		\kappa_{xyy}	=-\frac{v\lambda \pi k_{B}^{2}\left[ \Delta ^{4} \lambda ^{2}\left( \lambda ^{2} -1\right) +\Delta ^{2} \mu ^{2} -\mu ^{4}\right]}{48\hbar\left( \Delta ^{2} \lambda ^{2} +\mu ^{2}\right)^{5/2}},
\end{equation}
where $\lambda=w/v$ is dimensionless, determining the topology of the Fermi surface.
 It is noteworthy that our result contrasts with the findings reported in Ref \cite{Zhou2022}, which is proportional to $T^2$. This difference stems from the reason that the latter result is contributed by Berry curvature in the case that $\mathcal{T}$ is preserved but inversion symmetry $\mathcal{P}$ is broken.
\CIV

%%%%%%%%%%%%%%%%%%%%%%%%%%%%%%%%%%%%%%%%%%%%%%%%%%%%%%%%%%%%%%%%%%%%%%%%%%%%%%%%%%%%%%%%%%%%%%%%%%%%%%%%%%%%%%
\begin{figure}[t]
	\centering
	\includegraphics[width=1\columnwidth]{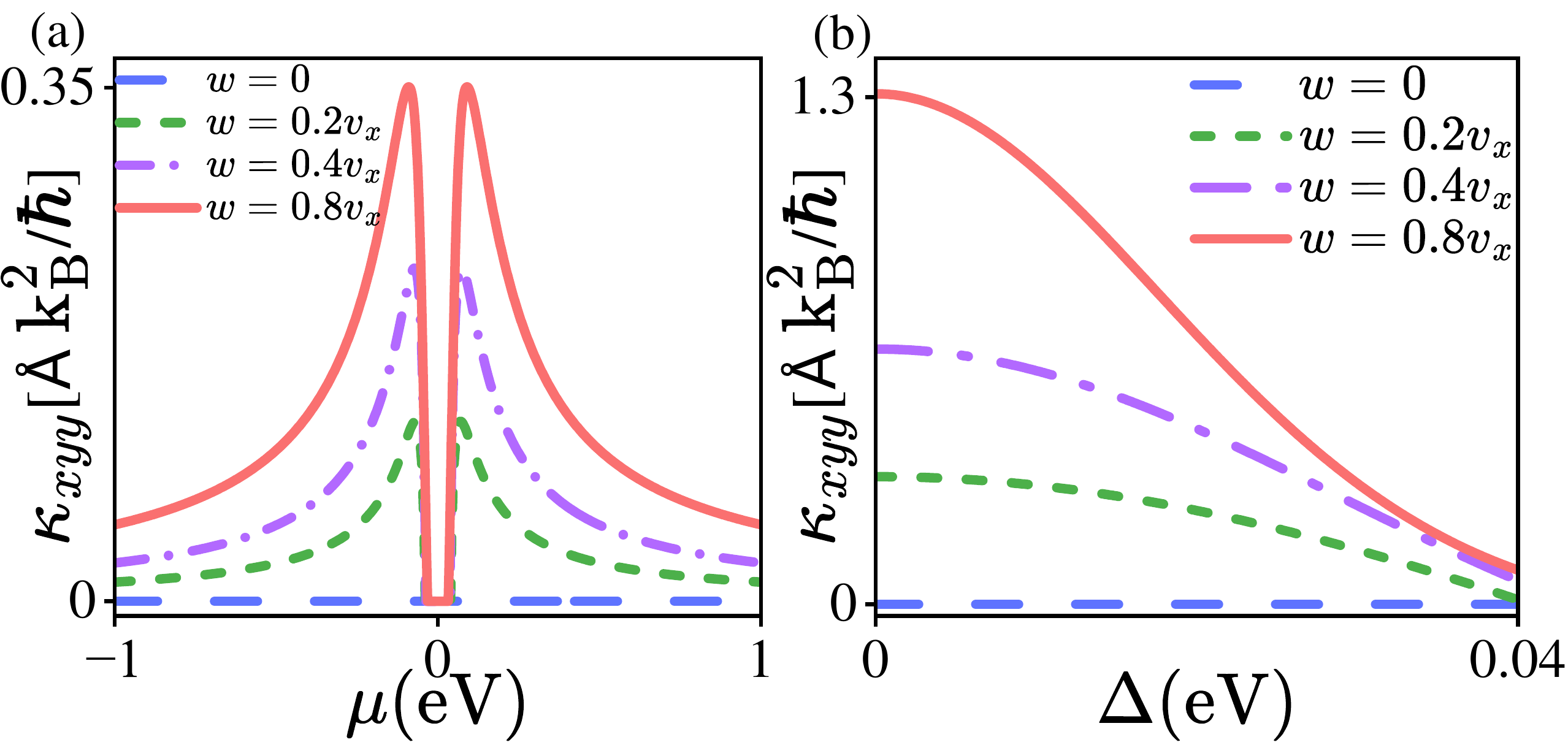}
	\caption{(a) The thermal conductivity $\kappa_{xyy}$ vs chemical potential $\mu$, we set $\Delta=40$ meV. (b) The thermal conductivity $\kappa_{xyy}$ vs $\Delta$, we set $\mu=40$ meV. The other parameters are set to $v_x=v_y=1$ eV$\mathring{\text{A}}$, $w/v_x=0$, $0.2$, $0.4$, $0.8$.}
	\label{kappa_Delta_T}
\end{figure}
%%%%%%%%%%%%%%%%%%%%%%%%%%%%%%%%%%%%%%%%%%%%%%%%%%%%%%%%%%%%%%%%%%%%%%%%%%%%%%%%%%%%%%%%%%%%%%%%%%%%%%%%%%%%%%

Notably, $\kappa_{xyy}$ is larger  due to large TBCP around the band edges. Fig. \ref{kappa_Delta_T}(a) illustrates the relationship %between intrinsic thermal conductivity
of $\kappa_{xyy}$  and  $\mu$.
Due to the ``Fermi surface" effect, $\kappa_{xyy}$ develops a peak rapidly and then decreases when $\mu$ enters into the edge of the band and gradually lifts up further.
It is worth mentioning that $\kappa_{xyy}$ is an even function of chemical potential $\mu$, which means the conduction and valence bands make the same intrinsic contribution.
Meanwhile, there is no the second-order thermal Hall conductivity in the absence of tilt term because of the inversion symmetry.
In other words, the tilt term is the key condition to ensure the violation of space inversion along the direction of $k_x$, resulting in the non-vanishing $\kappa_{xyy}$.
Furthermore, with the increase of $w$, $\kappa_{xyy}$ rises as the degree of inversion symmetry breaking increases.
%
%
%In passing, we also consider other parameters of the intrinsic thermal conductivity $\kappa_{xyy}$  [
In Fig. \ref{kappa_Delta_T}(b),  the dependence of $\kappa_{xyy}$ on $\Delta$ is shown.
 \CV As $\Delta$ increases to match the chemical potential, $\kappa_{xyy}$ gradually approaches zero but does not fully reach it, owing to the existence of  the small quantity $\mathcal{O}(\lambda^2)$ in Eq. (\ref{kappajiexi}).\CIV  %$\kappa_{xyy}$ exhibits excellent tunability for various parameters. It's always true that the larger the tilt $w$, the more obvious the change of $\kappa_{xyy}$.

%%%%%%%%%%%%%%%%%%%%%%%%%%%%%%%%%%%%%%%%%%%%%%%%%%%%%%%%%%%%%%%%%%%%%%%%%%%%%%%%%%%%%%%%%%%%%%%%%
\begin{figure}[t]
	\centering
	\includegraphics[width=1\columnwidth]{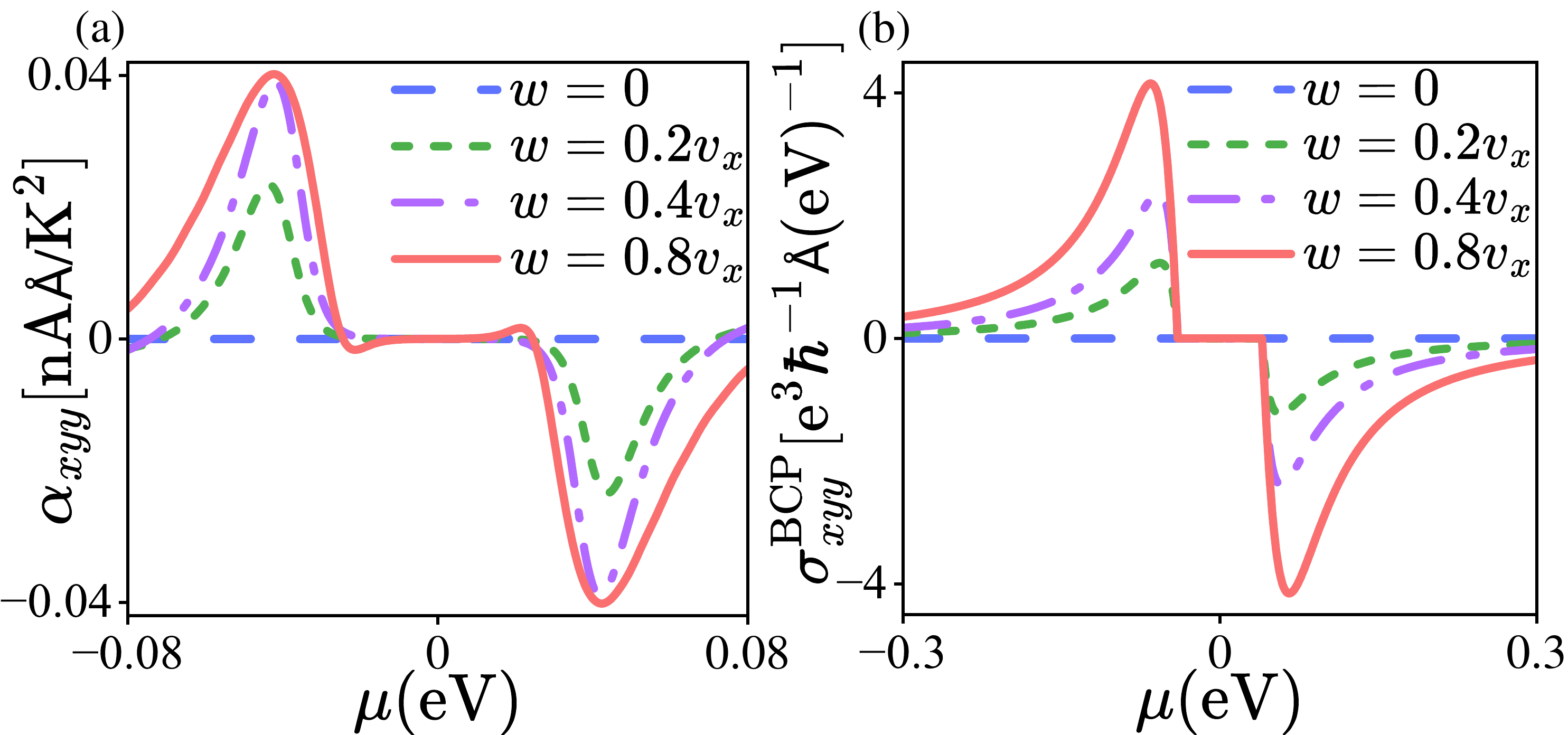}
	\caption{(a) Nernst coefficient $\alpha_{xyy}$ vs $\mu$, we set $k_BT=$ 2.6 meV ($T\sim$ 30 K). (b) Conductivity $\sigma_{xyy}^\text{BCP}$ vs $\mu$. The parameters are set to $v_x=v_y=1$ eV$\mathring{\text{A}}$, $\Delta=40$ meV. }
	\label{kappa_xyy_jiexi}
\end{figure}
%%%%%%%%%%%%%%%%%%%%%%%%%%%%%%%%%%%%%%%%%%%%%%%%%%%%%%%%%%%%%%%%%%%%%%%%%%%%%%%%%%%%%%%%%%%%%%%%%%%

The second-order Nernst coefficient $\alpha_{xyy}$ vs $\mu$ is numerically computed in Fig. \ref{kappa_xyy_jiexi}(a). %{\color{blue} It is seen, $L_{xyy}$ gradually increases from small number and then approaches a saturation level as $\mu$ gradually increases.} %due to its ``Fermi sea'' contribution.}
There are two peaks but with opposite signs around the edge of valence and conduction bands respectively.
Due to symmetry constraints, the BCD-induced $\sigma^{\text{BCD}}_{xyy}$ disappears under $\mathcal{PT}$ symmetry.  %In addition,
Drude-induced $\sigma^{\text{Drude}}_{xyy}$ is not considered, and we only consider BCP-induced $\sigma^{\text{BCP}}_{xyy}$. %second order Hall current (call it $\sigma_{abc}$ for short).
For a comparison, we plot the dependence of the second-order intrinsic conductivity $\sigma^{\text{BCP}}_{xyy}$ on $\mu$ in Fig. \ref{kappa_xyy_jiexi}(b), where $\sigma^{\text{BCP}}_{xyy}$ decreases to zero as $\mu$ moves deeper into the conduction or valence bands.
There are two peaks with opposite signs around the edges of valence bands and conduction bands, respectively.

With these calculations, it is reasonable to analyze the ratio relationship between the second-order intrinsic thermal conductivity \CV coefficient \CIV and the second-order intrinsic electrical conductivity \CV coefficient \CIV (represented as $\sigma_{xyy}$ for short).
We \CV can precisely obtain \CIV the second-order intrinsic Wiedemann-Franz law as
\CV
\begin{equation}\begin{aligned}\label{wflaw}
		\kappa _{xyy} =-\frac{L}{2e}\left( \mu -\frac{\Delta ^{2} \lambda ^{2}}{\mu }\right) \sigma _{xyy},
\end{aligned}\end{equation}
\CIV
where $L$ is the Lorentz number.
Clearly, the second-order intrinsic thermal conductivity \CV coefficient \CIV and second-order intrinsic conductivity \CV coefficient \CIV are directly proportional to each other, and is determined by $\mu$. \CV The observed behavior stems from the different distributions of the TBCP [Fig. \ref{G_yy}(a)-(c)], responsible for the thermal conductivity coefficient, and the BCP \cite{Liu2021}, which determines the electrical conductivity coefficient. This discrepancy in their distributions is likely the cause of the chemical potential $\mu$-dependent nature of the proposed WF law. \CIV
The second-order intrinsic WF law exhibits a direct proportionality between the second-order intrinsic thermal and electric transport, akin to the linear WF law.
It differs from the second-order WF law due to the BCD mechanism \CV($\sigma_{abc}$ $=$ $-\frac{e}{2L}\frac{\partial\kappa_{abc}}{\partial\mu}$)\CIV \cite{Wang2022}, and it is tentatively ascribed to the disparity between the intrinsic and extrinsic mechanisms of the BCP and the BCD mechanism.

%\CV With the second-order extrinsic \cite{Wang2022,Zeng2020} and intrinsic effect proposed here, our work augments the nonlinear theory of thermoelectric transport and offers valuable perspectives for experimental studies. The second-order intrinsic WF law we propose is effective for materials with nonzero (T)BCP, which is of great significance for the design and optimization of thermoelectric materials, such as thermoelectric generators and refrigerators. Moreover, by adjusting the electronic structure and chemical potential of materials, we can control their thermal and electrical conductivities, thereby optimizing their thermoelectric efficiency. This provides theoretical guidance for the design of new thermoelectric materials.\CIV

Until now the most of research had been focused on external contributions \cite{Sodemann2015,Du2018,Ma2018,Xiao2020}. Since the proposal of the intrinsic NAHE \cite{Gao2014}, the intrinsic nonlinear current has drawn increasing attention as it assumes a dominant role in $\mathcal{PT}$-symmetric materials where the external contributions vanish.
Recently, the impact of impurities and disorder in samples on transport effects has been explored \cite{Sinitsyn2005,Du2019,Atencia2023,Arai2024,Ding2019}. However, it remains uncertain whether the second-order intrinsic transport phenomenon in $\mathcal{PT}$-symmetric materials is associated with impurities or disorder.
%
%More recently, NbMnP \cite{Arai2024} has been employed to assess how impurities affect thermoelectric transport in AFM spin structures.
Owing to its $\mathcal{PT}$ symmetry, MnBi$_2$Te$_4$ \cite{Gao2023,Deng2020,Li2024a,He2020,Lin2022,Li2023} permits the presence of intrinsic NAHE, which has attracted extensive attention in recent years.
To acquire an understanding of the magnitude of the second-order intrinsic thermoelectric response, MnBi$_2$Te$_4$ could be an ideal candidate material for detecting the proposed intrinsic second-order WF law and Nernst effect. %transport currents.
%
%With subsequent in-depth research, in an effort to enhance the precision and applicability of the theoretical model, previously overlooked factors are progressively incorporated, and further modifications should be made to the theory.	
%
It should be emphasized that the proposed intrinsic second-order WF law links the transverse thermal and electric transport, thus is topologically dissipationless. Since the linear transverse WF law disappears, thus this intrinsic second-order transverse WF law is dominant. However, the longitudinal linear WF law still exists. Therefore, it leaves us an important open issue to explore the scaling relation between these two existing WF laws in the future.

\begin{acknowledgments}
This work is supported by the National Key R\&D Program of China (Grant No. 2024YFA1409200, No. 2022YFA1402802), CAS Project for Young Scientists in Basic Research Grant No. YSBR-057, the Strategic Priority Research Program of CAS (Grants No. XDB28000000, and No. XDB33000000), and the Training Program of Major Research plan of the National Natural Science Foundation of China (Grant No. 92165105).
\end{acknowledgments}

\begin{widetext}

\appendix

\section{The derivation of second-order intrinsic anomalous transport coefficients}\label{sec:A}
\subsection{The derivation of second-order intrinsic thermal conductivity coefficient}\label{app:k_abc_unsimply}
The second-order intrinsic thermal conductivity is $\kappa_{abc}$. 
By substituting the expressions $\tilde{\varepsilon}_{n,\boldsymbol{k}}$ and $\tilde{\boldsymbol{\Omega}}_n(\boldsymbol{k})$ into the nonlinear anomalous thermal current, $\boldsymbol{j}^{Q}_{{T}}$ can be derived into
\begin{equation}\begin{aligned}\boldsymbol{j}^{Q}_{{T}}&=- \frac{k_B^2T}{\hbar}\boldsymbol{\nabla}T\times\sum_m\int[d{\boldsymbol{k}}]\tilde{\boldsymbol{\Omega}}_m(\boldsymbol{k})\bigg[\frac{(\tilde{\varepsilon}_{m,\boldsymbol{k}}-\mu)^2}{(k_BT)^2}f^m_0+\frac{\pi^2}{3}-\ln^2(1-f_0^m)-2\mathrm{Li}_2(1-f_0^m)\bigg].\end{aligned}\end{equation}

In this paper, we mainly focus on the second-order transport phenomenon, which can be obtained by selectively collecting the terms proportional to $(\nabla T)^2$. Intrinsic second-order transverse thermal current in the direction $a$ is given by
\begin{equation}
	\begin{aligned}
		j_{{T},a}^{Q(2)}=&\kappa_{abc}\CV(-{\nabla_{b}T})(-{\nabla_{c}T})\CIV\\=&-\frac{k_{B}^{2}T}{\hbar}\sum_m\int[d{\boldsymbol{k}}]\left[ \boldsymbol\nabla T\times(\boldsymbol\nabla\times \boldsymbol{\mathcal{A}}_m^{(1)}(\boldsymbol{k}))\right] _{a} \bigg[\frac{(\varepsilon_{m,\boldsymbol{k}}-\mu)^{2}}{(k_{B}T)^{2}}f^m_0+\frac{\pi^2}{3}-\ln^2(1-f_0^m)-2\mathrm{Li}_2(1-f_0^m)\bigg]\\
		=&\frac{k_{B}^{2}}{\hbar}\epsilon_{ab}\sum_m\int[d{\boldsymbol{k}}]\partial_{a}\mathbb{G}^{\text{T}}_{m,bc}(\boldsymbol{k})\bigg[\frac{(\varepsilon_{m,\boldsymbol{k}}-\mu)^{2}}{(k_{B}T)^{2}}f^m_0+\frac{\pi^2}{3}-\ln^2(1-f_0^m)-2\mathrm{Li}_2(1-f_0^m)\bigg]\CV(-\nabla_{b}T)(-\nabla_{c}T)\CIV,
		\label{eq:thermal current}
	\end{aligned}
\end{equation}
where $\kappa_{abc}$ is second-order thermal conductivity and
$\CV\left\lbrace  \boldsymbol\nabla T\times[\boldsymbol\nabla\times \boldsymbol{\mathcal{A}}_m^{(1)}(\boldsymbol{k})]\right\rbrace  _{a}\CIV=-\frac{1}{T}\epsilon_{ab}\partial_{a}\mathbb{G}^{\text{T}}_{m,bc}(\boldsymbol{k})\nabla_{b}T\nabla_{c}T$ is used.

As a result, we can get second-order intrinsic thermal conductivity in Eq. (\ref{eq:thermal current}):
\begin{equation}\begin{aligned}\label{k_TBCP}
		\kappa_{abc}
		=&\frac{k_{B}^{2}}{\hbar}\epsilon_{ab}\sum_m\int[d{\boldsymbol{k}}]\partial_{a}\mathbb{G}^{\text{T}}_{m,bc}(\boldsymbol{k})\bigg[\frac{{(\varepsilon_{m,\boldsymbol{k}}-\mu)^{2}}f_{0}^m}{(k_{B}T)^{2}}+\frac{\pi^2}{3}-\ln^2(1-f_0^m)-2\mathrm{Li}_2(1-f_0^m)\bigg].
\end{aligned}\end{equation}

Considering the low temperature limit \cite{RoyKarmakar2022,Yokoyama2011}, \CV the first and third terms in square brackets cancel out in Eq. (\ref{k_TBCP}), thus \CIV
\begin{equation}\begin{aligned}\label{huajiankTBCP}
		\kappa_{abc}
		=&\frac{k_{B}^{2}}{\hbar}\epsilon_{ab}\sum_n\int[d{\boldsymbol{k}}]\partial_{a}\mathbb{G}^{\text{T}}_{n,bc}(\boldsymbol{k})\bigg[\frac{\pi^2}{3}-2\mathrm{Li}_2(1-f_0^n)\bigg]\\
		=&\frac{\pi^2}3\frac{k_{B}^{2}}\hbar\epsilon_{ab}\sum_n\int[d{\boldsymbol{k}}]\partial_{a}\mathbb{G}^{\text{T}}_{n,bc}(\boldsymbol{k})\theta(\mu-\varepsilon_{n,\boldsymbol{k}})\\
		=&\CV\frac{\pi^2}3\frac{k_{B}^{2}}\hbar\epsilon_{ab}\sum_n\int[d{\boldsymbol{k}}]\mathbb{G}^{\text{T}}_{n,bc}(\boldsymbol{k})\delta(\mu-\varepsilon_{n,\boldsymbol{k}})\frac{\partial \varepsilon_{n,\boldsymbol{k}}}{\partial k_a}\CIV\\
		=&\frac{\pi^2}3\frac{k_{B}^{3}T}\hbar\epsilon_{ab}P^{\text{T}}_{a,bc}(\mu),
\end{aligned}\end{equation}
\CV where $\theta(\mu-\varepsilon_{n,\boldsymbol{k}})$ is step Function,  $\text{Li}_2(0)=0$, $\text{Li}_2(1)=\pi^2/6$, \CIV  and we define $P^{\text{T}}_{a,bc}(\mu)$ as:
\begin{equation}
	\begin{aligned}P^{\text{T}}_{a,bc}(\mu)=\frac{1}{k_BT}\sum_n\int[d\boldsymbol{k}]\mathbb{G}^{\text{T}}_{n,bc}(\boldsymbol{k})\delta(\mu-\varepsilon_{n,\boldsymbol{k}})\frac{\partial\varepsilon_{n,\boldsymbol{k}}}{\partial k_{a}} .\end{aligned}\end{equation}	

By partial integration, the existence of $\partial f_0^n/\partial \varepsilon_{n, {\boldsymbol{k}}}$ shows that the intrinsic nonlinear thermal current is related to the ``Fermi-surface'' effect,
indicating that only the bands near the Fermi surface contribute to the integration, and is expected to be vanished in the band gap.

\subsection{The derivation of second-order intrinsic Nernst coefficient }\label{app:Derivation of L}
The second-order intrinsic Nernst coefficient is expressed by $\alpha_{abc}$. 
By substituting the expressions $\tilde{\varepsilon}_{n,\boldsymbol{k}}$ and $\tilde{\boldsymbol{\Omega}}_m(\boldsymbol{k})$ into the Nernst-like current, which is given by:
\begin{equation}\begin{aligned}\boldsymbol{j}^{N}=- \frac{e}{\hbar}\boldsymbol{\nabla}T\times\sum_m\int[d{\boldsymbol{k}}]\tilde{\boldsymbol{\Omega}}_m(\boldsymbol{k})\bigg[\frac{(\tilde{\varepsilon}_{m,\boldsymbol{k}}-\mu)}{T}f_0^{m}+k_B\ln(1+e^{-\beta(\tilde\varepsilon_{\boldsymbol{m,k}}-\mu)})\bigg],\end{aligned}\end{equation}
thus, second-order intrinsic Nernst-like current in the $a$ direction:
\begin{equation}
	\begin{aligned}
		J_{a}^{N,(2)}=&\alpha_{abc}\CV(-\nabla_{b}T) (-\nabla_{c}T)\CIV\\		=&-\frac{e}{\hbar}\sum_m\int[d{\boldsymbol{k}}]\bigg[\boldsymbol{\nabla} T\times(\boldsymbol{\nabla}\times \boldsymbol{\mathcal{A}}_m^{(1)}(\boldsymbol{k}))\bigg]_{a}\bigg[\frac{\varepsilon_{m,\boldsymbol{k}}-\mu}{T}f_{0}^m+k_B\ln(1+e^{-\beta(\varepsilon_{m,\boldsymbol{k}}-\mu)})\bigg]\\		
		=&\frac{e}{\hbar T}\sum_m\int[d{\boldsymbol{k}}]\epsilon_{ab}\partial_{a}\mathbb{G}^{\text{T}}_{m,bc}(\boldsymbol{k})S_{m}(\mathbf{r},\boldsymbol{k})\CV(-{\nabla_{b}T})(-{\nabla_{c}T})\CIV,
		\label{eq:Nernst current}
\end{aligned}\end{equation}
where $S_{m}(\mathbf{r},\boldsymbol{k})=\frac{(\varepsilon_{m,\boldsymbol{k}}-\mu)}Tf_0^m+k_B\ln\left(1+e^{-\beta(\varepsilon_{m,\boldsymbol{k}}-\mu)}\right) $ is entropy density.

Furthermore, we obtain the second-order intrinsic Nernst coefficient $\alpha_{abc}$ in Eq. (\ref{eq:Nernst current}), \begin{equation}\begin{aligned}	\label{L_TBCP}	\alpha_{abc}
		=&\frac{e}{\hbar T}\epsilon_{ab}\sum_m\int[d{\boldsymbol{k}}]\partial_{a}\mathbb{G}^{\text{T}}_{m,bc}(\boldsymbol{k})S_{m}(\mathbf{r},\boldsymbol{k}).
\end{aligned}\end{equation}

\subsection{The derivation of second-order intrinsic Hall conductivity coefficient}
The second-order intrinsic Hall conductivity reads $\sigma_{abc}$. 
The Berry curvature $\tilde{\boldsymbol{\Omega}}_n^{\text{E}}(\boldsymbol{k})$ is
corrected to first order in electric field $\mathbb E$: $\tilde{\boldsymbol{\Omega}}_n^{\text{E}}(\boldsymbol{k})={\boldsymbol{\Omega}}_n(\boldsymbol{k})+{\boldsymbol{\Omega}}^{\text{E}(1)}_n(\boldsymbol{k})$, ${\boldsymbol{\Omega}}_n(\boldsymbol{k})$ is the unperturbed Berry curvature, ${\boldsymbol{\Omega}}^{\text{E}(1)}_n(\boldsymbol{k})$ is the first-order correction to the Berry curvature ${\boldsymbol{\Omega}}^{\text{E}(1)}_n(\boldsymbol{k})=\boldsymbol{\nabla}_{\boldsymbol{k}}\times\boldsymbol{\mathcal{A}}_n^{\text{E}(1)}(\boldsymbol{k})$, where $\boldsymbol{\mathcal{A}}_n^{\text{E}(1)}(\boldsymbol{k})$ is the first-order correction of electric field $\mathbb E$ to the Berry connection, ${\mathcal{A}}_{n,a}^{\text{E}(1)}(\boldsymbol{k})=\mathbb{G}^\text{E}_{n,ab}(\boldsymbol{k})\mathbb{E}_b$, 
in which the BCP tensor \cite{Gao2014} is defined as 
\begin{align}
  \mathbb{G}_{n,ab}^{\text{E}}(\boldsymbol{k})=2e\Re\sum_{l\neq n}\frac{\mathcal{A}_{nl,a}(\boldsymbol{k})\mathcal{A}_{ln,b}(\boldsymbol{k})}{\varepsilon_{n,\boldsymbol{k}}-\varepsilon_{l,\boldsymbol{k}}}.
\end{align}

Employing the Boltzmann equation, $\dot{\boldsymbol{k}}\cdot\nabla_{\boldsymbol{k}}f_{\boldsymbol{k}}^n=-\frac{f_{\boldsymbol{k}}^n-f_0^n}{\tau}$,
the	non-equilibrium distribution function can be obtained as:
\begin{equation}f_{\boldsymbol{k}}^n=\sum_{m=0}^\infty\left( \frac{e\tau}{\hbar} \mathbf{E}\cdot\boldsymbol \nabla_{\boldsymbol k}\right) ^mf_0^n,\end{equation}

The $a$-direction component of the second-order current $ \boldsymbol{j}^{E,(2)}$ is denoted as $j^{E,(2)}_{a} =\sigma_{abc}E_bE_c$. Subsequently, the second-order conductivity can be divided into three parts \cite{Zhang2023,Kaplan2023,Gao2014,Liu2021,Wang2021,Sodemann2015}:
$\sigma_{abc}=\sigma^{\text{Drude}}_{abc}+\sigma^\text{BCP}_{abc}+\sigma^\text{BCD}_{abc}$,
with
\begin{align}\label{eq:sigma_abc_bcp}
	\sigma^\text{BCP}_{abc}=&-\frac{e^2}{\hbar}\epsilon_{ab}\sum_m\int[d{\boldsymbol{k}}]\partial _a\mathbb{G}^{\text{E}}_{m,bc}(\boldsymbol{k})f_{0}^m.
\end{align}

In this context, we have only analyzed the second-order electrical conductivity contributed by BCP, utilizing the Sommerfeld expansion \cite{Mermin1976},
\CV	\begin{equation}\begin{aligned}\label{eq: sommerfeld expansion}
		\int_{-\infty}^{+\infty}d\varepsilon\frac{\partial f_0^n}{\partial\varepsilon}\Phi(\varepsilon)=-\Phi(\mu)-\frac{\pi^{2}}{6}k_B^{2}T^2\left[ \frac{\partial^2 \Phi(\varepsilon)}{\partial \varepsilon^2}\right]\Bigg| _{\varepsilon=\mu}-\frac{7\pi^{4}}{360}k_B^{4}T^{4}\left[ \frac{\partial^4 \Phi(\varepsilon)}{\partial \varepsilon^4}\right]\Bigg|_{\varepsilon=\mu},
\end{aligned}\end{equation}
\CIV
we can simplify Eq. (\ref{eq:sigma_abc_bcp}) to Eq. (\ref{sigma_abc}),
\begin{equation}\begin{aligned}\label{sigma_abc}
		\sigma^\text{BCP}_{abc}=&-\frac{e^2}{\hbar}\epsilon_{ab}\sum_n\int[d{\boldsymbol{k}}]\partial _a\mathbb{G}^{\text{E}}_{n,bc}(\boldsymbol{k})f_{0}^n \\
		&=\frac{e^3}{\hbar}\epsilon_{ab}\int d\varepsilon\left\{\frac{1}{e}\int\sum_n[d\boldsymbol{k}]\mathbb{G}_{n,bc}^{{\text{E}}}(\boldsymbol{k})\delta(\varepsilon-\varepsilon_{n,\boldsymbol{k}})\frac{\partial\varepsilon_{n,\boldsymbol{k}}}{\partial k_{a}}\right\}\frac{\partial f_{0}^n}{\partial\varepsilon}\\
		&=-\frac{e^3}{\hbar}\epsilon_{ab}\left[ P_{a,bc}^{{\text{E}}}(\mu)+\frac{\pi^{2}}{6}(k_{B}T)^{2}P_{a,bc}^{{\text{E}}{(2)}}(\mu)\right] ,
\end{aligned}\end{equation}
where, $P_{a,bc}^{{\text{E}}(n)}(\mu)=\partial^nP_{a,bc}^{{\text{E}}}(\mu)/\partial\mu^n$, and $P^{\text{E}}_{a,bc}(\mu)$ is defined as
\begin{equation}\begin{aligned}
		P^{\text{E}}_{a,bc}(\mu)=\frac{1}{e}\int\sum_n[d\boldsymbol{k}]\mathbb{G}_{n,bc}^{\text{E}}\boldsymbol(\boldsymbol{k})\delta(\mu-\varepsilon_{n,\boldsymbol{k}})\frac{\partial\varepsilon_{n,\boldsymbol{k}}}{\partial k_{a}}.
\end{aligned}\end{equation}
\CV\section{The derivation of the second-order intrinsic  Wiedemann-Franz law}\CIV\label{APP:WFLAW}
\CV
From Eqs. (\ref{huajiankTBCP}) and (\ref{Gexpression}), we can obtain the second-order intrinsic thermal conductivity coefficient for the valence band,
\begin{equation}\begin{aligned}\label{huajiank_TBCP}
		\kappa_{xyy}
		=&\frac{\pi^2}3\frac{k_{B}^{3}T}\hbar\epsilon_{xy}P^{\text{T}}_{x,yy}(\mu)\\
		=&\frac{\pi ^{2}}{3}\frac{k_{B}^{3} T}{\hbar }\frac{1}{k_{B} T}\int [d\boldsymbol k]\left[ \mathbb{G}_{1,yy}^{\text{T}}(\boldsymbol k)\frac{\partial \varepsilon _{1,\boldsymbol k}}{\partial k_{x}} -\mathbb{G}_{1,xy}^{\text{T}}(\boldsymbol k)\frac{\partial \varepsilon _{1,\boldsymbol k}}{\partial k_{y}}\right] \delta (\varepsilon _{1,\boldsymbol k} -\mu )\\
		=&\frac{\pi ^{2}}{3}\frac{k_{B}^{3} T}{\hbar }\frac{1}{( 2\pi )^{2} k_{B} T}\int\int dk_{x} dk_{y}\left[-\frac{w^{2} k_{x} v^{2}}{4E^{5}_{0} (\boldsymbol k)} (\Delta ^{2} +k_{x}^{2} v^{2} )+\frac{wv^{4} k_{x}^{2} E_{0}( \boldsymbol k)}{4E^{5}_{0} (\boldsymbol k)}\right] \delta [ wk_{x} -E_{0}(\boldsymbol k) \ -\mu ].
\end{aligned}\end{equation}

We can incorporate the $v$ factor into $k_x$ and $k_y$  by redefining  $k_x$  as $k_x/v$  and $k_y$ as $k_y/v$, providing a more convenient form for subsequent calculations and analysis. Then from the $\delta [ wk_{x} -E_{0}(\boldsymbol k) \ -\mu ]$ function in Eq. (\ref{huajiank_TBCP}), we can get
\begin{equation}k_{y} =\pm \sqrt{( \mu -\lambda k_{x})^{2} -\Delta ^{2} -k_{x}^{2}},\end{equation} and the range of \(k_x\) is given by \(k_x\in [k_a,k_b]\) on the Fermi surface, where the values of \(k_a\) and \(k_b\) are expressed as \(k_{a/b}=\frac{-\mu\lambda\mp\sqrt{\mu^{2}-\Delta^{2}+\lambda^{2}\Delta^{2}}}{1 - \lambda^{2}}\).

Therefore, the second-order intrinsic thermal conductivity coefficient $\kappa_{xyy}$ for the conduction bands in Eq. (\ref{huajiank_TBCP}) by using the method in~\cite{Liu2021} is written as :
\begin{equation}\begin{aligned}\label{jiexik_TBCP}
		\kappa_{xyy}
			=&\frac{\pi ^{2}}{3}\frac{k_{B}^{2}}{\hbar }\frac{v \lambda}{( 2\pi )^{2}}\int_{k_a}^{k_b} dk_{x}\left[\frac{k_{x}^{2}( \lambda k_{x} -\mu ) -\lambda k_{x} (\Delta ^{2} +k_{x}^{2} )}{4( \lambda k_{x} -\mu )^{5}}\right]\frac{2( \lambda k_{x} -\mu )}{| k_{y}| }\Bigg| _{k_{y} =\pm \sqrt{( \mu -\lambda k_{x})^{2} -\Delta ^{2} -k_{x}^{2}}}\\
			=&\frac{\pi ^{2}}{3}\frac{k_{B}^{2}}{\hbar }\frac{v \lambda}{8\pi ^{2}}\int_{k_a}^{k_b} dk_{x}\frac{k_{x}^{2}( \lambda k_{x} -\mu ) -\lambda k_{x} (\Delta ^{2} +k_{x}^{2} )}{( \lambda k_{x} -\mu )^{4}}\frac{1}{\sqrt{( \mu -\lambda k_{x})^{2} -\Delta ^{2} -k_{x}^{2}}}\\
			=&-\frac{v\lambda \pi k_{B}^{2}}{48\hbar }\frac{\left[ \Delta ^{4} \lambda ^{2}\left( \lambda ^{2} -1\right) +\Delta ^{2} \mu ^{2} -\mu ^{4}\right]}{\left( \Delta ^{2} \lambda ^{2} +\mu ^{2}\right)^{5/2}},
\end{aligned}\end{equation}
where $\lambda=w/v$, quantifying the magnitude of band tilting.

For  the second-order intrinsic Hall conductivity coefficient $\sigma_{xyy}$, only the influence of BCP mechanism on the coefficient is considered, the analytical expression of $\sigma_{xyy}$ is given in Ref. \cite{Liu2021}, which is

\begin{equation}\begin{aligned}\label{sigma_BCP}
	\sigma_{xyy}=-\frac{e^{3}v\lambda \mu \left[ \Delta ^{2}\left( \lambda ^{2} -1\right) +\mu ^{2}\right]}{8\pi \hbar\left( \Delta ^{2} \lambda ^{2} +\mu ^{2}\right)^{5/2}},
\end{aligned}\end{equation}
 and for the conduction bands, the identical result in Eqs. (\ref{huajiank_TBCP})  and (\ref{sigma_BCP}) can be achieved.

Therefore, by synthesizing the results obtained above, we can successfully derive the second-order intrinsic  Wiedemann-Franz law,
\begin{equation}\begin{aligned}
\kappa _{xyy} =-\frac{L}{2e}\left( \mu -\frac{\Delta ^{2} \lambda ^{2}}{\mu }\right) \sigma _{xyy}
\end{aligned}\end{equation}

\CIV
\end{widetext}
%\onecolumngrid

%\newpage

%\bibliography{test}
%\bibliography{reference}

\begin{thebibliography}{67}%
\makeatletter
\providecommand \@ifxundefined [1]{%
	\@ifx{#1\undefined}
}%
\providecommand \@ifnum [1]{%
	\ifnum #1\expandafter \@firstoftwo
	\else \expandafter \@secondoftwo
	\fi
}%
\providecommand \@ifx [1]{%
	\ifx #1\expandafter \@firstoftwo
	\else \expandafter \@secondoftwo
	\fi
}%
\providecommand \natexlab [1]{#1}%
\providecommand \enquote  [1]{``#1''}%
\providecommand \bibnamefont  [1]{#1}%
\providecommand \bibfnamefont [1]{#1}%
\providecommand \citenamefont [1]{#1}%
\providecommand \href@noop [0]{\@secondoftwo}%
\providecommand \href [0]{\begingroup \@sanitize@url \@href}%
\providecommand \@href[1]{\@@startlink{#1}\@@href}%
\providecommand \@@href[1]{\endgroup#1\@@endlink}%
\providecommand \@sanitize@url [0]{\catcode `\\12\catcode `\$12\catcode
	`\&12\catcode `\#12\catcode `\^12\catcode `\_12\catcode `\%12\relax}%
\providecommand \@@startlink[1]{}%
\providecommand \@@endlink[0]{}%
\providecommand \url  [0]{\begingroup\@sanitize@url \@url }%
\providecommand \@url [1]{\endgroup\@href {#1}{\urlprefix }}%
\providecommand \urlprefix  [0]{URL }%
\providecommand \Eprint [0]{\href }%
\providecommand \doibase [0]{https://doi.org/}%
\providecommand \selectlanguage [0]{\@gobble}%
\providecommand \bibinfo  [0]{\@secondoftwo}%
\providecommand \bibfield  [0]{\@secondoftwo}%
\providecommand \translation [1]{[#1]}%
\providecommand \BibitemOpen [0]{}%
\providecommand \bibitemStop [0]{}%
\providecommand \bibitemNoStop [0]{.\EOS\space}%
\providecommand \EOS [0]{\spacefactor3000\relax}%
\providecommand \BibitemShut  [1]{\csname bibitem#1\endcsname}%
\let\auto@bib@innerbib\@empty
%</preamble>
\bibitem [{\citenamefont {Shen}(2017)}]{Shen2017}%
\BibitemOpen
\bibfield  {author} {\bibinfo {author} {\bibfnamefont {S.-Q.}\ \bibnamefont
		{Shen}},\ }\href {https://doi.org/10.1007/978-981-10-4606-3} { {\bibinfo
		{title} {Topological Insulators}}}\ (\bibinfo  {publisher} {Springer
	Singapore},\ \bibinfo {year} {2017})\BibitemShut {NoStop}%
\bibitem [{\citenamefont {Xiao}\ \emph {et~al.}(2010)\citenamefont {Xiao},
	\citenamefont {Chang},\ and\ \citenamefont {Niu}}]{Xiao2010}%
\BibitemOpen
\bibfield  {author} {\bibinfo {author} {\bibfnamefont {D.}~\bibnamefont
		{Xiao}}, \bibinfo {author} {\bibfnamefont {M.-C.}\ \bibnamefont {Chang}},\
	and\ \bibinfo {author} {\bibfnamefont {Q.}~\bibnamefont {Niu}},\ }\bibfield
{title} {\bibinfo {title} {Berry phase effects on electronic properties},\
}\href {https://doi.org/10.1103/revmodphys.82.1959} {\bibfield  {journal}
	{\bibinfo  {journal} {Rev. Mod. Phys.}\ }\textbf {\bibinfo {volume} {82}},\
	\bibinfo {pages} {1959} (\bibinfo {year} {2010})}\BibitemShut {NoStop}%
\bibitem [{\citenamefont {Tsirkin}\ and\ \citenamefont
	{Souza}(2022)}]{Tsirkin2022}%
\BibitemOpen
\bibfield  {author} {\bibinfo {author} {\bibfnamefont {S.~S.}\ \bibnamefont
		{Tsirkin}}\ and\ \bibinfo {author} {\bibfnamefont {I.}~\bibnamefont
		{Souza}},\ }\bibfield  {title} {\bibinfo {title} {{On the separation of Hall
			and Ohmic nonlinear responses}},\ }\href
{https://doi.org/10.21468/SciPostPhysCore.5.3.039} {\bibfield  {journal}
	{\bibinfo  {journal} {SciPost Phys. Core}\ }\textbf {\bibinfo {volume} {5}},\
	\bibinfo {pages} {039} (\bibinfo {year} {2022})}\BibitemShut {NoStop}%
\bibitem [{\citenamefont {Wang}\ \emph {et~al.}(2024)\citenamefont {Wang},
	\citenamefont {Zhang}, \citenamefont {Zhu},\ and\ \citenamefont
	{Su}}]{Wang2024}%
\BibitemOpen
\bibfield  {author} {\bibinfo {author} {\bibfnamefont {Y.}~\bibnamefont
		{Wang}}, \bibinfo {author} {\bibfnamefont {Z.}~\bibnamefont {Zhang}},
	\bibinfo {author} {\bibfnamefont {Z.-G.}\ \bibnamefont {Zhu}},\ and\ \bibinfo
	{author} {\bibfnamefont {G.}~\bibnamefont {Su}},\ }\bibfield  {title}
{\bibinfo {title} {Intrinsic nonlinear ohmic current},\ }\href
{https://doi.org/10.1103/physrevb.109.085419} {\bibfield  {journal} {\bibinfo
		{journal} {Phys. Rev. B}\ }\textbf {\bibinfo {volume} {109}},\ \bibinfo
	{pages} {085419} (\bibinfo {year} {2024})}\BibitemShut {NoStop}%
\bibitem [{\citenamefont {Kaplan}\ \emph {et~al.}(2024)\citenamefont {Kaplan},
	\citenamefont {Holder},\ and\ \citenamefont {Yan}}]{Kaplan2024}%
\BibitemOpen
\bibfield  {author} {\bibinfo {author} {\bibfnamefont {D.}~\bibnamefont
		{Kaplan}}, \bibinfo {author} {\bibfnamefont {T.}~\bibnamefont {Holder}},\
	and\ \bibinfo {author} {\bibfnamefont {B.}~\bibnamefont {Yan}},\ }\bibfield
{title} {\bibinfo {title} {Unification of nonlinear anomalous Hall effect and
		nonreciprocal magnetoresistance in metals by the quantum geometry},\ }\href
{https://doi.org/10.1103/physrevlett.132.026301} {\bibfield  {journal}
	{\bibinfo  {journal} {Phys. Rev. Lett.}\ }\textbf {\bibinfo {volume} {132}},\
	\bibinfo {pages} {026301} (\bibinfo {year} {2024})}\BibitemShut {NoStop}%
\bibitem [{\citenamefont {Jia}\ \emph {et~al.}(2024)\citenamefont {Jia},
	\citenamefont {Xiang}, \citenamefont {Qiao},\ and\ \citenamefont
	{Wang}}]{Jia2024}%
\BibitemOpen
\bibfield  {author} {\bibinfo {author} {\bibfnamefont {J.}~\bibnamefont
		{Jia}}, \bibinfo {author} {\bibfnamefont {L.}~\bibnamefont {Xiang}}, \bibinfo
	{author} {\bibfnamefont {Z.}~\bibnamefont {Qiao}},\ and\ \bibinfo {author}
	{\bibfnamefont {J.}~\bibnamefont {Wang}},\ }\bibfield  {title} {\bibinfo
	{title} {On the equivalence of the semiclassical theory and the response
		theory},\ }\href@noop {} {\  (\bibinfo {year} {2024})},\ \Eprint
{https://arxiv.org/abs/2404.17086} {arXiv:2404.17086 [cond-mat.mes-hall]}
\BibitemShut {NoStop}%
\bibitem [{\citenamefont {Nag}\ \emph {et~al.}(2023)\citenamefont {Nag},
	\citenamefont {Das}, \citenamefont {Zeng},\ and\ \citenamefont
	{Nandy}}]{Nag2023}%
\BibitemOpen
\bibfield  {author} {\bibinfo {author} {\bibfnamefont {T.}~\bibnamefont
		{Nag}}, \bibinfo {author} {\bibfnamefont {S.~K.}\ \bibnamefont {Das}},
	\bibinfo {author} {\bibfnamefont {C.}~\bibnamefont {Zeng}},\ and\ \bibinfo
	{author} {\bibfnamefont {S.}~\bibnamefont {Nandy}},\ }\bibfield  {title}
{\bibinfo {title} {Third-order hall effect in the surface states of a
		topological insulator},\ }\href {https://doi.org/10.1103/physrevb.107.245141}
{\bibfield  {journal} {\bibinfo  {journal} {Phys. Rev. B}\ }\textbf {\bibinfo
		{volume} {107}},\ \bibinfo {pages} {245141} (\bibinfo {year}
	{2023})}\BibitemShut {NoStop}%
\bibitem [{\citenamefont {Sodemann}\ and\ \citenamefont
	{Fu}(2015)}]{Sodemann2015}%
\BibitemOpen
\bibfield  {author} {\bibinfo {author} {\bibfnamefont {I.}~\bibnamefont
		{Sodemann}}\ and\ \bibinfo {author} {\bibfnamefont {L.}~\bibnamefont {Fu}},\
}\bibfield  {title} {\bibinfo {title} {Quantum nonlinear Hall effect induced
		by Berry curvature dipole in time-reversal invariant materials},\ }\href
{https://doi.org/10.1103/physrevlett.115.216806} {\bibfield  {journal}
	{\bibinfo  {journal} {Phys. Rev. Lett.}\ }\textbf {\bibinfo {volume} {115}},\
	\bibinfo {pages} {216806} (\bibinfo {year} {2015})}\BibitemShut {NoStop}%
\bibitem [{\citenamefont {Du}\ \emph {et~al.}(2018)\citenamefont {Du},
	\citenamefont {Wang}, \citenamefont {Lu},\ and\ \citenamefont
	{Xie}}]{Du2018}%
\BibitemOpen
\bibfield  {author} {\bibinfo {author} {\bibfnamefont {Z.}~\bibnamefont
		{Du}}, \bibinfo {author} {\bibfnamefont {C.}~\bibnamefont {Wang}}, \bibinfo
	{author} {\bibfnamefont {H.-Z.}\ \bibnamefont {Lu}},\ and\ \bibinfo {author}
	{\bibfnamefont {X.}~\bibnamefont {Xie}},\ }\bibfield  {title} {\bibinfo
	{title} {Band signatures for strong nonlinear Hall effect in bilayer WTe$_2$},\
}\href {https://doi.org/10.1103/physrevlett.121.266601} {\bibfield  {journal}
	{\bibinfo  {journal} {Phys. Rev. Lett.}\ }\textbf {\bibinfo {volume} {121}},\
	\bibinfo {pages} {266601} (\bibinfo {year} {2018})}\BibitemShut {NoStop}%
\bibitem [{\citenamefont {Ma}\ \emph {et~al.}(2018)\citenamefont {Ma},
	\citenamefont {Xu}, \citenamefont {Shen}, \citenamefont {MacNeill},
	\citenamefont {Fatemi}, \citenamefont {Chang}, \citenamefont {Mier~Valdivia},
	\citenamefont {Wu}, \citenamefont {Du}, \citenamefont {Hsu}, \citenamefont
	{Fang}, \citenamefont {Gibson}, \citenamefont {Watanabe}, \citenamefont
	{Taniguchi}, \citenamefont {Cava}, \citenamefont {Kaxiras}, \citenamefont
	{Lu}, \citenamefont {Lin}, \citenamefont {Fu}, \citenamefont {Gedik},\ and\
	\citenamefont {Jarillo-Herrero}}]{Ma2018}%
\BibitemOpen
\bibfield  {author} {\bibinfo {author} {\bibfnamefont {Q.}~\bibnamefont
		{Ma}}, \bibinfo {author} {\bibfnamefont {S.-Y.}\ \bibnamefont {Xu}}, \bibinfo
	{author} {\bibfnamefont {H.}~\bibnamefont {Shen}}, \bibinfo {author}
	{\bibfnamefont {D.}~\bibnamefont {MacNeill}}, \bibinfo {author}
	{\bibfnamefont {V.}~\bibnamefont {Fatemi}}, \bibinfo {author} {\bibfnamefont
		{T.-R.}\ \bibnamefont {Chang}}, \bibinfo {author} {\bibfnamefont {A.~M.}\
		\bibnamefont {Mier~Valdivia}}, \bibinfo {author} {\bibfnamefont
		{S.}~\bibnamefont {Wu}}, \bibinfo {author} {\bibfnamefont {Z.}~\bibnamefont
		{Du}}, \bibinfo {author} {\bibfnamefont {C.-H.}\ \bibnamefont {Hsu}},
	\bibinfo {author} {\bibfnamefont {S.}~\bibnamefont {Fang}}, \bibinfo {author}
	{\bibfnamefont {Q.~D.}\ \bibnamefont {Gibson}}, \bibinfo {author}
	{\bibfnamefont {K.}~\bibnamefont {Watanabe}}, \bibinfo {author}
	{\bibfnamefont {T.}~\bibnamefont {Taniguchi}}, \bibinfo {author}
	{\bibfnamefont {R.~J.}\ \bibnamefont {Cava}}, \bibinfo {author}
	{\bibfnamefont {E.}~\bibnamefont {Kaxiras}}, \bibinfo {author} {\bibfnamefont
		{H.-Z.}\ \bibnamefont {Lu}}, \bibinfo {author} {\bibfnamefont
		{H.}~\bibnamefont {Lin}}, \bibinfo {author} {\bibfnamefont {L.}~\bibnamefont
		{Fu}}, \bibinfo {author} {\bibfnamefont {N.}~\bibnamefont {Gedik}},\ and\
	\bibinfo {author} {\bibfnamefont {P.}~\bibnamefont {Jarillo-Herrero}},\
}\bibfield  {title} {\bibinfo {title} {Observation of the nonlinear Hall
		effect under time-reversal-symmetric conditions},\ }\href
{https://doi.org/10.1038/s41586-018-0807-6} {\bibfield  {journal} {\bibinfo
		{journal} {Nature}\ }\textbf {\bibinfo {volume} {565}},\ \bibinfo {pages}
	{337} (\bibinfo {year} {2018})}\BibitemShut {NoStop}%
\bibitem [{\citenamefont {Xiao}\ \emph {et~al.}(2020)\citenamefont {Xiao},
	\citenamefont {Shao}, \citenamefont {Huang},\ and\ \citenamefont
	{Jiang}}]{Xiao2020}%
\BibitemOpen
\bibfield  {author} {\bibinfo {author} {\bibfnamefont {R.-C.}\ \bibnamefont
		{Xiao}}, \bibinfo {author} {\bibfnamefont {D.-F.}\ \bibnamefont {Shao}},
	\bibinfo {author} {\bibfnamefont {W.}~\bibnamefont {Huang}},\ and\ \bibinfo
	{author} {\bibfnamefont {H.}~\bibnamefont {Jiang}},\ }\bibfield  {title}
{\bibinfo {title} {Electrical detection of ferroelectriclike metals through
		the nonlinear Hall effect},\ }\href
{https://doi.org/10.1103/physrevb.102.024109} {\bibfield  {journal} {\bibinfo
		{journal} {Phys. Rev. B}\ }\textbf {\bibinfo {volume} {102}},\ \bibinfo
	{pages} {024109} (\bibinfo {year} {2020})}\BibitemShut {NoStop}%
\bibitem [{\citenamefont {Zeng}\ \emph {et~al.}(2021)\citenamefont {Zeng},
	\citenamefont {Nandy},\ and\ \citenamefont {Tewari}}]{Zeng2021}%
\BibitemOpen
\bibfield  {author} {\bibinfo {author} {\bibfnamefont {C.}~\bibnamefont
		{Zeng}}, \bibinfo {author} {\bibfnamefont {S.}~\bibnamefont {Nandy}},\ and\
	\bibinfo {author} {\bibfnamefont {S.}~\bibnamefont {Tewari}},\ }\bibfield
{title} {\bibinfo {title} {Nonlinear transport in Weyl semimetals induced by
		Berry curvature dipole},\ }\href
{https://doi.org/10.1103/physrevb.103.245119} {\bibfield  {journal} {\bibinfo
		{journal} {Phys. Rev. B}\ }\textbf {\bibinfo {volume} {103}},\ \bibinfo
	{pages} {245119} (\bibinfo {year} {2021})}\BibitemShut {NoStop}%
\bibitem [{\citenamefont {Zhang}\ \emph {et~al.}(2018)\citenamefont {Zhang},
	\citenamefont {Sun},\ and\ \citenamefont {Yan}}]{Zhang2018}%
\BibitemOpen
\bibfield  {author} {\bibinfo {author} {\bibfnamefont {Y.}~\bibnamefont
		{Zhang}}, \bibinfo {author} {\bibfnamefont {Y.}~\bibnamefont {Sun}},\ and\
	\bibinfo {author} {\bibfnamefont {B.}~\bibnamefont {Yan}},\ }\bibfield
{title} {\bibinfo {title} {Berry curvature dipole in Weyl semimetal
		materials: An ab initio study},\ }\href
{https://doi.org/10.1103/physrevb.97.041101} {\bibfield  {journal} {\bibinfo
		{journal} {Phys. Rev. B}\ }\textbf {\bibinfo {volume} {97}},\ \bibinfo
	{pages} {041101} (\bibinfo {year} {2018})}\BibitemShut {NoStop}%
\bibitem [{\citenamefont {Wang}\ \emph
	{et~al.}(2022{\natexlab{a}})\citenamefont {Wang}, \citenamefont {Xiao},
	\citenamefont {Liu}, \citenamefont {Zhang}, \citenamefont {Lai},
	\citenamefont {Zhu}, \citenamefont {Cai}, \citenamefont {Wang}, \citenamefont
	{Chen}, \citenamefont {Deng}, \citenamefont {Liu}, \citenamefont {Yang},\
	and\ \citenamefont {Gao}}]{Wang2022a}%
\BibitemOpen
\bibfield  {author} {\bibinfo {author} {\bibfnamefont {C.}~\bibnamefont
		{Wang}}, \bibinfo {author} {\bibfnamefont {R.-C.}\ \bibnamefont {Xiao}},
	\bibinfo {author} {\bibfnamefont {H.}~\bibnamefont {Liu}}, \bibinfo {author}
	{\bibfnamefont {Z.}~\bibnamefont {Zhang}}, \bibinfo {author} {\bibfnamefont
		{S.}~\bibnamefont {Lai}}, \bibinfo {author} {\bibfnamefont {C.}~\bibnamefont
		{Zhu}}, \bibinfo {author} {\bibfnamefont {H.}~\bibnamefont {Cai}}, \bibinfo
	{author} {\bibfnamefont {N.}~\bibnamefont {Wang}}, \bibinfo {author}
	{\bibfnamefont {S.}~\bibnamefont {Chen}}, \bibinfo {author} {\bibfnamefont
		{Y.}~\bibnamefont {Deng}}, \bibinfo {author} {\bibfnamefont {Z.}~\bibnamefont
		{Liu}}, \bibinfo {author} {\bibfnamefont {S.~A.}\ \bibnamefont {Yang}},\ and\
	\bibinfo {author} {\bibfnamefont {W.-B.}\ \bibnamefont {Gao}},\ }\bibfield
{title} {\bibinfo {title} {Room-temperature third-order nonlinear Hall effect
		in Weyl semimetal TaIrTe$_4$},\ }\href {https://doi.org/10.1093/nsr/nwac020}
{\bibfield  {journal} {\bibinfo  {journal} {Natl. Sci. Rev.}\ }\textbf
	{\bibinfo {volume} {9}},\ \bibinfo {pages} {nwac020} (\bibinfo {year}
	{2022}{\natexlab{a}})}\BibitemShut {NoStop}%
\bibitem [{\citenamefont {Zhang}\ \emph {et~al.}(2022)\citenamefont {Zhang},
	\citenamefont {Xiao}, \citenamefont {Zhou}, \citenamefont {Hu}, \citenamefont
	{Xie}, \citenamefont {Yan},\ and\ \citenamefont {Law}}]{Zhang2022}%
\BibitemOpen
\bibfield  {author} {\bibinfo {author} {\bibfnamefont {C.-P.}\ \bibnamefont
		{Zhang}}, \bibinfo {author} {\bibfnamefont {J.}~\bibnamefont {Xiao}},
	\bibinfo {author} {\bibfnamefont {B.~T.}\ \bibnamefont {Zhou}}, \bibinfo
	{author} {\bibfnamefont {J.-X.}\ \bibnamefont {Hu}}, \bibinfo {author}
	{\bibfnamefont {Y.-M.}\ \bibnamefont {Xie}}, \bibinfo {author} {\bibfnamefont
		{B.}~\bibnamefont {Yan}},\ and\ \bibinfo {author} {\bibfnamefont {K.~T.}\
		\bibnamefont {Law}},\ }\bibfield  {title} {\bibinfo {title} {Giant nonlinear
		Hall effect in strained twisted bilayer graphene},\ }\href
{https://doi.org/10.1103/physrevb.106.l041111} {\bibfield  {journal}
	{\bibinfo  {journal} {Phys. Rev. B}\ }\textbf {\bibinfo {volume} {106}},\
	\bibinfo {pages} {L041111} (\bibinfo {year} {2022})}\BibitemShut {NoStop}%
\bibitem [{\citenamefont {Akatsuka}\ \emph {et~al.}(2024)\citenamefont
	{Akatsuka}, \citenamefont {Sakano}, \citenamefont {Yamamoto}, \citenamefont
	{Nomoto}, \citenamefont {Arita}, \citenamefont {Murata}, \citenamefont
	{Sasagawa}, \citenamefont {Watanabe}, \citenamefont {Taniguchi},
	\citenamefont {Mitsuishi}, \citenamefont {Kitamura}, \citenamefont {Horiba},
	\citenamefont {Sugawara}, \citenamefont {Souma}, \citenamefont {Sato},
	\citenamefont {Kumigashira}, \citenamefont {Shinokita}, \citenamefont {Wang},
	\citenamefont {Matsuda}, \citenamefont {Masubuchi}, \citenamefont {Machida},\
	and\ \citenamefont {Ishizaka}}]{Akatsuka2024}%
\BibitemOpen
\bibfield  {author} {\bibinfo {author} {\bibfnamefont {S.}~\bibnamefont
		{Akatsuka}}, \bibinfo {author} {\bibfnamefont {M.}~\bibnamefont {Sakano}},
	\bibinfo {author} {\bibfnamefont {T.}~\bibnamefont {Yamamoto}}, \bibinfo
	{author} {\bibfnamefont {T.}~\bibnamefont {Nomoto}}, \bibinfo {author}
	{\bibfnamefont {R.}~\bibnamefont {Arita}}, \bibinfo {author} {\bibfnamefont
		{R.}~\bibnamefont {Murata}}, \bibinfo {author} {\bibfnamefont
		{T.}~\bibnamefont {Sasagawa}}, \bibinfo {author} {\bibfnamefont
		{K.}~\bibnamefont {Watanabe}}, \bibinfo {author} {\bibfnamefont
		{T.}~\bibnamefont {Taniguchi}}, \bibinfo {author} {\bibfnamefont
		{N.}~\bibnamefont {Mitsuishi}}, \bibinfo {author} {\bibfnamefont
		{M.}~\bibnamefont {Kitamura}}, \bibinfo {author} {\bibfnamefont
		{K.}~\bibnamefont {Horiba}}, \bibinfo {author} {\bibfnamefont
		{K.}~\bibnamefont {Sugawara}}, \bibinfo {author} {\bibfnamefont
		{S.}~\bibnamefont {Souma}}, \bibinfo {author} {\bibfnamefont
		{T.}~\bibnamefont {Sato}}, \bibinfo {author} {\bibfnamefont {H.}~\bibnamefont
		{Kumigashira}}, \bibinfo {author} {\bibfnamefont {K.}~\bibnamefont
		{Shinokita}}, \bibinfo {author} {\bibfnamefont {H.}~\bibnamefont {Wang}},
	\bibinfo {author} {\bibfnamefont {K.}~\bibnamefont {Matsuda}}, \bibinfo
	{author} {\bibfnamefont {S.}~\bibnamefont {Masubuchi}}, \bibinfo {author}
	{\bibfnamefont {T.}~\bibnamefont {Machida}},\ and\ \bibinfo {author}
	{\bibfnamefont {K.}~\bibnamefont {Ishizaka}},\ }\bibfield  {title} {\bibinfo
	{title} {180°-twisted bilayer ReSe$_2$ as an artificial noncentrosymmetric
		semiconductor},\ }\href {https://doi.org/10.1103/physrevresearch.6.l022048}
{\bibfield  {journal} {\bibinfo  {journal} {Phys. Rev. Research}\ }\textbf
	{\bibinfo {volume} {6}},\ \bibinfo {pages} {l022048} (\bibinfo {year}
	{2024})}\BibitemShut {NoStop}%
\bibitem [{\citenamefont {Pang}\ \emph {et~al.}(2024)\citenamefont {Pang},
	\citenamefont {Jin},\ and\ \citenamefont {He}}]{Pang2024}%
\BibitemOpen
\bibfield  {author} {\bibinfo {author} {\bibfnamefont {H.}~\bibnamefont
		{Pang}}, \bibinfo {author} {\bibfnamefont {G.}~\bibnamefont {Jin}},\ and\
	\bibinfo {author} {\bibfnamefont {L.}~\bibnamefont {He}},\ }\bibfield
{title} {\bibinfo {title} {Tuning of Berry-curvature dipole in taas slabs: An
		effective route to enhance the nonlinear Hall response},\ }\href
{https://doi.org/10.1103/physrevmaterials.8.043403} {\bibfield  {journal}
	{\bibinfo  {journal} {Phys. Rev. Materials}\ }\textbf {\bibinfo {volume}
		{8}},\ \bibinfo {pages} {043403} (\bibinfo {year} {2024})}\BibitemShut
{NoStop}%
\bibitem [{\citenamefont {Gao}\ \emph {et~al.}(2014)\citenamefont {Gao},
	\citenamefont {Yang},\ and\ \citenamefont {Niu}}]{Gao2014}%
\BibitemOpen
\bibfield  {author} {\bibinfo {author} {\bibfnamefont {Y.}~\bibnamefont
		{Gao}}, \bibinfo {author} {\bibfnamefont {S.~A.}\ \bibnamefont {Yang}},\ and\
	\bibinfo {author} {\bibfnamefont {Q.}~\bibnamefont {Niu}},\ }\bibfield
{title} {\bibinfo {title} {Field induced positional shift of bloch electrons
		and its dynamical implications},\ }\href
{https://doi.org/10.1103/physrevlett.112.166601} {\bibfield  {journal}
	{\bibinfo  {journal} {Phys. Rev. Lett.}\ }\textbf {\bibinfo {volume} {112}},\
	\bibinfo {pages} {166601} (\bibinfo {year} {2014})}\BibitemShut {NoStop}%
\bibitem [{\citenamefont {Gao}\ \emph {et~al.}(2015)\citenamefont {Gao},
	\citenamefont {Yang},\ and\ \citenamefont {Niu}}]{Gao2015}%
\BibitemOpen
\bibfield  {author} {\bibinfo {author} {\bibfnamefont {Y.}~\bibnamefont
		{Gao}}, \bibinfo {author} {\bibfnamefont {S.~A.}\ \bibnamefont {Yang}},\ and\
	\bibinfo {author} {\bibfnamefont {Q.}~\bibnamefont {Niu}},\ }\bibfield
{title} {\bibinfo {title} {Geometrical effects in orbital magnetic
		susceptibility},\ }\href {https://doi.org/10.1103/physrevb.91.214405}
{\bibfield  {journal} {\bibinfo  {journal} {Phys. Rev. B}\ }\textbf {\bibinfo
		{volume} {91}},\ \bibinfo {pages} {214405} (\bibinfo {year}
	{2015})}\BibitemShut {NoStop}%
\bibitem [{\citenamefont {Liu}\ \emph {et~al.}(2021)\citenamefont {Liu},
	\citenamefont {Zhao}, \citenamefont {Huang}, \citenamefont {Wu},
	\citenamefont {Sheng}, \citenamefont {Xiao},\ and\ \citenamefont
	{Yang}}]{Liu2021}%
\BibitemOpen
\bibfield  {author} {\bibinfo {author} {\bibfnamefont {H.}~\bibnamefont
		{Liu}}, \bibinfo {author} {\bibfnamefont {J.}~\bibnamefont {Zhao}}, \bibinfo
	{author} {\bibfnamefont {Y.-X.}\ \bibnamefont {Huang}}, \bibinfo {author}
	{\bibfnamefont {W.}~\bibnamefont {Wu}}, \bibinfo {author} {\bibfnamefont
		{X.-L.}\ \bibnamefont {Sheng}}, \bibinfo {author} {\bibfnamefont
		{C.}~\bibnamefont {Xiao}},\ and\ \bibinfo {author} {\bibfnamefont {S.~A.}\
		\bibnamefont {Yang}},\ }\bibfield  {title} {\bibinfo {title} {Intrinsic
		second-order anomalous Hall effect and its application in compensated
		antiferromagnets},\ }\href {https://doi.org/10.1103/physrevlett.127.277202}
{\bibfield  {journal} {\bibinfo  {journal} {Phys. Rev. Lett.}\ }\textbf
	{\bibinfo {volume} {127}},\ \bibinfo {pages} {277202} (\bibinfo {year}
	{2021})}\BibitemShut {NoStop}%
\bibitem [{\citenamefont {Wang}\ \emph {et~al.}(2021)\citenamefont {Wang},
	\citenamefont {Gao},\ and\ \citenamefont {Xiao}}]{Wang2021}%
\BibitemOpen
\bibfield  {author} {\bibinfo {author} {\bibfnamefont {C.}~\bibnamefont
		{Wang}}, \bibinfo {author} {\bibfnamefont {Y.}~\bibnamefont {Gao}},\ and\
	\bibinfo {author} {\bibfnamefont {D.}~\bibnamefont {Xiao}},\ }\bibfield
{title} {\bibinfo {title} {Intrinsic nonlinear Hall effect in
		antiferromagnetic tetragonal CuMnAs},\ }\href
{https://doi.org/10.1103/physrevlett.127.277201} {\bibfield  {journal}
	{\bibinfo  {journal} {Phys. Rev. Lett.}\ }\textbf {\bibinfo {volume} {127}},\
	\bibinfo {pages} {277201} (\bibinfo {year} {2021})}\BibitemShut {NoStop}%
\bibitem [{\citenamefont {Liu}\ \emph {et~al.}(2022)\citenamefont {Liu},
	\citenamefont {Zhao}, \citenamefont {Huang}, \citenamefont {Feng},
	\citenamefont {Xiao}, \citenamefont {Wu}, \citenamefont {Lai}, \citenamefont
	{Gao},\ and\ \citenamefont {Yang}}]{Liu2022}%
\BibitemOpen
\bibfield  {author} {\bibinfo {author} {\bibfnamefont {H.}~\bibnamefont
		{Liu}}, \bibinfo {author} {\bibfnamefont {J.}~\bibnamefont {Zhao}}, \bibinfo
	{author} {\bibfnamefont {Y.-X.}\ \bibnamefont {Huang}}, \bibinfo {author}
	{\bibfnamefont {X.}~\bibnamefont {Feng}}, \bibinfo {author} {\bibfnamefont
		{C.}~\bibnamefont {Xiao}}, \bibinfo {author} {\bibfnamefont {W.}~\bibnamefont
		{Wu}}, \bibinfo {author} {\bibfnamefont {S.}~\bibnamefont {Lai}}, \bibinfo
	{author} {\bibfnamefont {W.-b.}\ \bibnamefont {Gao}},\ and\ \bibinfo {author}
	{\bibfnamefont {S.~A.}\ \bibnamefont {Yang}},\ }\bibfield  {title} {\bibinfo
	{title} {Berry connection polarizability tensor and third-order Hall
		effect},\ }\href {https://doi.org/10.1103/physrevb.105.045118} {\bibfield
	{journal} {\bibinfo  {journal} {Phys. Rev. B}\ }\textbf {\bibinfo {volume}
		{105}},\ \bibinfo {pages} {045118} (\bibinfo {year} {2022})}\BibitemShut
{NoStop}%
\bibitem [{\citenamefont {Pal}\ and\ \citenamefont {Ghosh}(2024)}]{Pal2024}%
\BibitemOpen
\bibfield  {author} {\bibinfo {author} {\bibfnamefont {O.}~\bibnamefont
		{Pal}}\ and\ \bibinfo {author} {\bibfnamefont {T.~K.}\ \bibnamefont
		{Ghosh}},\ }\bibfield  {title} {\bibinfo {title} {Polarization and
		third-order Hall effect in $\text{\uppercase\expandafter{\romannumeral3}}$-V semiconductor heterojunctions},\ }\href
{https://doi.org/10.1103/physrevb.109.035202} {\bibfield  {journal} {\bibinfo
		{journal} {Phys. Rev. B}\ }\textbf {\bibinfo {volume} {109}},\ \bibinfo
	{pages} {035202} (\bibinfo {year} {2024})}\BibitemShut {NoStop}%
\bibitem [{\citenamefont {Zhang}\ \emph {et~al.}(2023)\citenamefont {Zhang},
	\citenamefont {Zhu},\ and\ \citenamefont {Su}}]{Zhang2023}%
\BibitemOpen
\bibfield  {author} {\bibinfo {author} {\bibfnamefont {Z.-F.}\ \bibnamefont
		{Zhang}}, \bibinfo {author} {\bibfnamefont {Z.-G.}\ \bibnamefont {Zhu}},\
	and\ \bibinfo {author} {\bibfnamefont {G.}~\bibnamefont {Su}},\ }\bibfield
{title} {\bibinfo {title} {Symmetry dictionary on charge and spin nonlinear
		responses for all magnetic point groups with nontrivial topological nature},\
}\href {https://doi.org/10.1093/nsr/nwad104} {\bibfield  {journal} {\bibinfo
		{journal} {Natl. Sci. Rev.}\ }\textbf {\bibinfo {volume} {10}},\ \bibinfo
	{pages} {nwad104} (\bibinfo {year} {2023})}\BibitemShut {NoStop}%
\bibitem [{\citenamefont {Li}\ and\ \citenamefont {Zhu}(2024)}]{Li2024}%
\BibitemOpen
\bibfield  {author} {\bibinfo {author} {\bibfnamefont {J.-C.}\ \bibnamefont
		{Li}}\ and\ \bibinfo {author} {\bibfnamefont {Z.-G.}\ \bibnamefont {Zhu}},\
}\bibfield  {title} {\bibinfo {title} {Intrinsic second-order magnon thermal
		Hall effect},\ }\href {https://doi.org/10.1088/1361-648x/ad5bb0} {\bibfield
	{journal} {\bibinfo  {journal} {J. Phys. Condens. Matter}\ }\textbf {\bibinfo
		{volume} {36}},\ \bibinfo {pages} {395802} (\bibinfo {year}
	{2024})}\BibitemShut {NoStop}%
\bibitem [{\citenamefont {Zhou}\ \emph {et~al.}(2022)\citenamefont {Zhou},
	\citenamefont {Zhang}, \citenamefont {Yu}, \citenamefont {Zhu},\ and\
	\citenamefont {Su}}]{Zhou2022}%
\BibitemOpen
\bibfield  {author} {\bibinfo {author} {\bibfnamefont {D.-K.}\ \bibnamefont
		{Zhou}}, \bibinfo {author} {\bibfnamefont {Z.-F.}\ \bibnamefont {Zhang}},
	\bibinfo {author} {\bibfnamefont {X.-Q.}\ \bibnamefont {Yu}}, \bibinfo
	{author} {\bibfnamefont {Z.-G.}\ \bibnamefont {Zhu}},\ and\ \bibinfo {author}
	{\bibfnamefont {G.}~\bibnamefont {Su}},\ }\bibfield  {title} {\bibinfo
	{title} {Fundamental distinction between intrinsic and extrinsic nonlinear
		thermal Hall effects},\ }\href {https://doi.org/10.1103/physrevb.105.l201103}
{\bibfield  {journal} {\bibinfo  {journal} {Phys. Rev. B}\ }\textbf {\bibinfo
		{volume} {105}},\ \bibinfo {pages} {l201103} (\bibinfo {year}
	{2022})}\BibitemShut {NoStop}%
\bibitem [{\citenamefont {Yu}\ \emph {et~al.}(2021)\citenamefont {Yu},
	\citenamefont {Zhu},\ and\ \citenamefont {Su}}]{Yu2021}%
\BibitemOpen
\bibfield  {author} {\bibinfo {author} {\bibfnamefont {X.-Q.}\ \bibnamefont
		{Yu}}, \bibinfo {author} {\bibfnamefont {Z.-G.}\ \bibnamefont {Zhu}},\ and\
	\bibinfo {author} {\bibfnamefont {G.}~\bibnamefont {Su}},\ }\bibfield
{title} {\bibinfo {title} {Hexagonal warping induced nonlinear planar Nernst
		effect in nonmagnetic topological insulators},\ }\href
{https://doi.org/10.1103/physrevb.103.035410} {\bibfield  {journal} {\bibinfo
		{journal} {Phys. Rev. B}\ }\textbf {\bibinfo {volume} {103}},\ \bibinfo
	{pages} {035410} (\bibinfo {year} {2021})}\BibitemShut {NoStop}%
\bibitem [{\citenamefont {Saha}\ and\ \citenamefont
	{Chowdhury}(2017)}]{Saha2017}%
\BibitemOpen
\bibfield  {author} {\bibinfo {author} {\bibfnamefont {M.}~\bibnamefont
		{Saha}}\ and\ \bibinfo {author} {\bibfnamefont {D.}~\bibnamefont
		{Chowdhury}},\ }\bibfield  {title} {\bibinfo {title} {Anomalous
		thermoelectric properties of a Floquet topological insulator with spin
		momentum non-orthogonality},\ }\href {https://doi.org/10.1063/1.4990969}
{\bibfield  {journal} {\bibinfo  {journal} {J. Appl. Phys.}\ }\textbf
	{\bibinfo {volume} {122}},\ \bibinfo {pages} {174301} (\bibinfo {year}
	{2017})}\BibitemShut {NoStop}%
\bibitem [{\citenamefont {Choudhari}\ and\ \citenamefont
	{Deo}(2019)}]{Choudhari2019}%
\BibitemOpen
\bibfield  {author} {\bibinfo {author} {\bibfnamefont {T.}~\bibnamefont
		{Choudhari}}\ and\ \bibinfo {author} {\bibfnamefont {N.}~\bibnamefont
		{Deo}},\ }\bibfield  {title} {\bibinfo {title} {Effect of hexagonal warping
		of the Fermi surface on the thermoelectric properties of a topological
		insulator irradiated with linearly polarized radiation},\ }\href
{https://doi.org/10.1103/physrevb.100.035303} {\bibfield  {journal} {\bibinfo
		{journal} {Phys. Rev. B}\ }\textbf {\bibinfo {volume} {100}},\ \bibinfo
	{pages} {035303} (\bibinfo {year} {2019})}\BibitemShut {NoStop}%
\bibitem [{\citenamefont {Roy~Karmakar}\ \emph {et~al.}(2022)\citenamefont
	{Roy~Karmakar}, \citenamefont {Nandy}, \citenamefont {Taraphder},\ and\
	\citenamefont {Das}}]{RoyKarmakar2022}%
\BibitemOpen
\bibfield  {author} {\bibinfo {author} {\bibfnamefont {A.}~\bibnamefont
		{Roy~Karmakar}}, \bibinfo {author} {\bibfnamefont {S.}~\bibnamefont {Nandy}},
	\bibinfo {author} {\bibfnamefont {A.}~\bibnamefont {Taraphder}},\ and\
	\bibinfo {author} {\bibfnamefont {G.~P.}\ \bibnamefont {Das}},\ }\bibfield
{title} {\bibinfo {title} {Giant anomalous thermal Hall effect in tilted
		type-$\text{\uppercase\expandafter{\romannumeral1}}$ magnetic Weyl semimetal Co$_3$Sn$_2$S$_2$},\ }\href
{https://doi.org/10.1103/physrevb.106.245133} {\bibfield  {journal} {\bibinfo
		{journal} {Phys. Rev. B}\ }\textbf {\bibinfo {volume} {106}},\ \bibinfo
	{pages} {245133} (\bibinfo {year} {2022})}\BibitemShut {NoStop}%
\bibitem [{\citenamefont {Nandy}\ \emph {et~al.}(2019)\citenamefont {Nandy},
	\citenamefont {Taraphder},\ and\ \citenamefont {Tewari}}]{Nandy2019}%
\BibitemOpen
\bibfield  {author} {\bibinfo {author} {\bibfnamefont {S.}~\bibnamefont
		{Nandy}}, \bibinfo {author} {\bibfnamefont {A.}~\bibnamefont {Taraphder}},\
	and\ \bibinfo {author} {\bibfnamefont {S.}~\bibnamefont {Tewari}},\
}\bibfield  {title} {\bibinfo {title} {Planar thermal Hall effect in Weyl
		semimetals},\ }\href {https://doi.org/10.1103/physrevb.100.115139} {\bibfield
	{journal} {\bibinfo  {journal} {Phys. Rev. B}\ }\textbf {\bibinfo {volume}
		{100}},\ \bibinfo {pages} {115139} (\bibinfo {year} {2019})}\BibitemShut
{NoStop}%
\bibitem [{\citenamefont {McCormick}\ \emph {et~al.}(2017)\citenamefont
	{McCormick}, \citenamefont {McKay},\ and\ \citenamefont
	{Trivedi}}]{McCormick2017}%
\BibitemOpen
\bibfield  {author} {\bibinfo {author} {\bibfnamefont {T.~M.}\ \bibnamefont
		{McCormick}}, \bibinfo {author} {\bibfnamefont {R.~C.}\ \bibnamefont
		{McKay}},\ and\ \bibinfo {author} {\bibfnamefont {N.}~\bibnamefont
		{Trivedi}},\ }\bibfield  {title} {\bibinfo {title} {Semiclassical theory of
		anomalous transport in type-$\text{\uppercase\expandafter{\romannumeral2}}$ topological Weyl semimetals},\ }\href
{https://doi.org/10.1103/physrevb.96.235116} {\bibfield  {journal} {\bibinfo
		{journal} {Phys. Rev. B}\ }\textbf {\bibinfo {volume} {96}},\ \bibinfo
	{pages} {235116} (\bibinfo {year} {2017})}\BibitemShut {NoStop}%
\bibitem [{\citenamefont {Bhalla}(2021)}]{Bhalla2021}%
\BibitemOpen
\bibfield  {author} {\bibinfo {author} {\bibfnamefont {P.}~\bibnamefont
		{Bhalla}},\ }\bibfield  {title} {\bibinfo {title} {Intrinsic contribution to
		nonlinear thermoelectric effects in topological insulators},\ }\href
{https://doi.org/10.1103/physrevb.103.115304} {\bibfield  {journal} {\bibinfo
		{journal} {Phys. Rev. B}\ }\textbf {\bibinfo {volume} {103}},\ \bibinfo
	{pages} {115304} (\bibinfo {year} {2021})}\BibitemShut {NoStop}%
\bibitem [{\citenamefont {Wang}\ \emph {et~al.}(2023)\citenamefont {Wang},
	\citenamefont {Zhu},\ and\ \citenamefont {Su}}]{Wang2023}%
\BibitemOpen
\bibfield  {author} {\bibinfo {author} {\bibfnamefont {Y.}~\bibnamefont
		{Wang}}, \bibinfo {author} {\bibfnamefont {Z.-G.}\ \bibnamefont {Zhu}},\ and\
	\bibinfo {author} {\bibfnamefont {G.}~\bibnamefont {Su}},\ }\bibfield
{title} {\bibinfo {title} {Field-induced Berry connection and anomalous
		planar Hall effect in tilted Weyl semimetals},\ }\href
{https://doi.org/10.1103/physrevresearch.5.043156} {\bibfield  {journal}
	{\bibinfo  {journal} {Phys. Rev. Research}\ }\textbf {\bibinfo {volume}
		{5}},\ \bibinfo {pages} {043156} (\bibinfo {year} {2023})}\BibitemShut
{NoStop}%
\bibitem [{\citenamefont {Qiang}\ \emph {et~al.}(2023)\citenamefont {Qiang},
	\citenamefont {Du}, \citenamefont {Lu},\ and\ \citenamefont
	{Xie}}]{Qiang2023}%
\BibitemOpen
\bibfield  {author} {\bibinfo {author} {\bibfnamefont {X.-B.}\ \bibnamefont
		{Qiang}}, \bibinfo {author} {\bibfnamefont {Z.~Z.}\ \bibnamefont {Du}},
	\bibinfo {author} {\bibfnamefont {H.-Z.}\ \bibnamefont {Lu}},\ and\ \bibinfo
	{author} {\bibfnamefont {X.~C.}\ \bibnamefont {Xie}},\ }\bibfield  {title}
{\bibinfo {title} {Topological and disorder corrections to the transverse
		Wiedemann-Franz law and Mott relation in kagome magnets and Dirac
		materials},\ }\href {https://doi.org/10.1103/physrevb.107.l161302} {\bibfield
	{journal} {\bibinfo  {journal} {Phys. Rev. B}\ }\textbf {\bibinfo {volume}
		{107}},\ \bibinfo {pages} {l161302} (\bibinfo {year} {2023})}\BibitemShut
{NoStop}%
\bibitem [{\citenamefont {Zeng}\ \emph {et~al.}(2020)\citenamefont {Zeng},
	\citenamefont {Nandy},\ and\ \citenamefont {Tewari}}]{Zeng2020}%
\BibitemOpen
\bibfield  {author} {\bibinfo {author} {\bibfnamefont {C.}~\bibnamefont
		{Zeng}}, \bibinfo {author} {\bibfnamefont {S.}~\bibnamefont {Nandy}},\ and\
	\bibinfo {author} {\bibfnamefont {S.}~\bibnamefont {Tewari}},\ }\bibfield
{title} {\bibinfo {title} {Fundamental relations for anomalous thermoelectric
		transport coefficients in the nonlinear regime},\ }\href
{https://doi.org/10.1103/physrevresearch.2.032066} {\bibfield  {journal}
	{\bibinfo  {journal} {Phys. Rev. Research}\ }\textbf {\bibinfo {volume}
		{2}},\ \bibinfo {pages} {032066} (\bibinfo {year} {2020})}\BibitemShut
{NoStop}%
\bibitem [{\citenamefont {Nakata}\ \emph {et~al.}(2022)\citenamefont {Nakata},
	\citenamefont {Ohnuma},\ and\ \citenamefont {Kim}}]{Nakata2022}%
\BibitemOpen
\bibfield  {author} {\bibinfo {author} {\bibfnamefont {K.}~\bibnamefont
		{Nakata}}, \bibinfo {author} {\bibfnamefont {Y.}~\bibnamefont {Ohnuma}},\
	and\ \bibinfo {author} {\bibfnamefont {S.~K.}\ \bibnamefont {Kim}},\
}\bibfield  {title} {\bibinfo {title} {Violation of the magnonic
		Wiedemann-Franz law in the strong nonlinear regime},\ }\href
{https://doi.org/10.1103/physrevb.105.184409} {\bibfield  {journal} {\bibinfo
		{journal} {Phys. Rev. B}\ }\textbf {\bibinfo {volume} {105}},\ \bibinfo
	{pages} {184409} (\bibinfo {year} {2022})}\BibitemShut {NoStop}%
\bibitem [{\citenamefont {Wang}\ \emph
	{et~al.}(2022{\natexlab{b}})\citenamefont {Wang}, \citenamefont {Zhu},\ and\
	\citenamefont {Su}}]{Wang2022}%
\BibitemOpen
\bibfield  {author} {\bibinfo {author} {\bibfnamefont {Y.}~\bibnamefont
		{Wang}}, \bibinfo {author} {\bibfnamefont {Z.-G.}\ \bibnamefont {Zhu}},\ and\
	\bibinfo {author} {\bibfnamefont {G.}~\bibnamefont {Su}},\ }\bibfield
{title} {\bibinfo {title} {Quantum theory of nonlinear thermal response},\
}\href {https://doi.org/10.1103/physrevb.106.035148} {\bibfield  {journal}
	{\bibinfo  {journal} {Phys. Rev. B}\ }\textbf {\bibinfo {volume} {106}},\
	\bibinfo {pages} {035148} (\bibinfo {year} {2022}{\natexlab{b}})}\BibitemShut
{NoStop}%
\bibitem [{\citenamefont {Neil~Ashcroft}(1976)}]{Mermin1976}%
\BibitemOpen
\bibfield  {author} {\bibinfo {author} {\bibfnamefont {N. W. Ashcroft, J. A. Mermin, and N. David},\ }\ }\bibinfo {title} {Solid State Physics}\ (\bibinfo
{publisher} {Brooks Cole; New edition (2 Jan. 1976)},\ \bibinfo {year}
{1976})\BibitemShut {NoStop}%
\bibitem [{\citenamefont {Smrcka}\ and\ \citenamefont
	{Streda}(1977)}]{Smrcka1977}%
\BibitemOpen
\bibfield  {author} {\bibinfo {author} {\bibfnamefont {L.}~\bibnamefont
		{Smrcka}}\ and\ \bibinfo {author} {\bibfnamefont {P.}~\bibnamefont
		{Streda}},\ }\bibfield  {title} {\bibinfo {title} {Transport coefficients in
		strong magnetic fields},\ }\href
{https://doi.org/10.1088/0022-3719/10/12/021} {\bibfield  {journal} {\bibinfo
		{journal} {J. Phys. C: Solid State Phys.}\ }\textbf {\bibinfo {volume}
		{10}},\ \bibinfo {pages} {2153} (\bibinfo {year} {1977})}\BibitemShut
{NoStop}%
\bibitem [{\citenamefont {Du}\ \emph {et~al.}(2021)\citenamefont {Du},
	\citenamefont {Lu},\ and\ \citenamefont {Xie}}]{Du2021}%
\BibitemOpen
\bibfield  {author} {\bibinfo {author} {\bibfnamefont {Z.~Z.}\ \bibnamefont
		{Du}}, \bibinfo {author} {\bibfnamefont {H.-Z.}\ \bibnamefont {Lu}},\ and\
	\bibinfo {author} {\bibfnamefont {X.~C.}\ \bibnamefont {Xie}},\ }\bibfield
{title} {\bibinfo {title} {Nonlinear Hall effects},\ }\href
{https://doi.org/10.1038/s42254-021-00359-6} {\bibfield  {journal} {\bibinfo
		{journal} {Nat. Rev. Phys.}\ }\textbf {\bibinfo {volume} {3}},\
	\bibinfo {pages} {744} (\bibinfo {year} {2021})}\BibitemShut {NoStop}%
\bibitem [{\citenamefont {Yokoyama}\ and\ \citenamefont
	{Murakami}(2011)}]{Yokoyama2011}%
\BibitemOpen
\bibfield  {author} {\bibinfo {author} {\bibfnamefont {T.}~\bibnamefont
		{Yokoyama}}\ and\ \bibinfo {author} {\bibfnamefont {S.}~\bibnamefont
		{Murakami}},\ }\bibfield  {title} {\bibinfo {title} {Transverse magnetic heat
		transport on the surface of a topological insulator},\ }\href
{https://doi.org/10.1103/physrevb.83.161407} {\bibfield  {journal} {\bibinfo
		{journal} {Phys. Rev. B}\ }\textbf {\bibinfo {volume} {83}},\ \bibinfo
	{pages} {161407} (\bibinfo {year} {2011})}\BibitemShut {NoStop}%
\bibitem [{\citenamefont {Bergman}\ and\ \citenamefont
	{Oganesyan}(2010)}]{Bergman2010}%
\BibitemOpen
\bibfield  {author} {\bibinfo {author} {\bibfnamefont {D.~L.}\ \bibnamefont
		{Bergman}}\ and\ \bibinfo {author} {\bibfnamefont {V.}~\bibnamefont
		{Oganesyan}},\ }\bibfield  {title} {\bibinfo {title} {Theory of
		dissipationless Nernst effects},\ }\href
{https://doi.org/10.1103/physrevlett.104.066601} {\bibfield  {journal}
	{\bibinfo  {journal} {Phys. Rev. Lett.}\ }\textbf {\bibinfo {volume} {104}},\
	\bibinfo {pages} {066601} (\bibinfo {year} {2010})}\BibitemShut {NoStop}%
\bibitem [{\citenamefont {Xiao}\ \emph {et~al.}(2006)\citenamefont {Xiao},
	\citenamefont {Yao}, \citenamefont {Fang},\ and\ \citenamefont
	{Niu}}]{Xiao2006}%
\BibitemOpen
\bibfield  {author} {\bibinfo {author} {\bibfnamefont {D.}~\bibnamefont
		{Xiao}}, \bibinfo {author} {\bibfnamefont {Y.}~\bibnamefont {Yao}}, \bibinfo
	{author} {\bibfnamefont {Z.}~\bibnamefont {Fang}},\ and\ \bibinfo {author}
	{\bibfnamefont {Q.}~\bibnamefont {Niu}},\ }\bibfield  {title} {\bibinfo
	{title} {Berry-phase effect in anomalous thermoelectric transport},\ }\href
{https://doi.org/10.1103/physrevlett.97.026603} {\bibfield  {journal}
	{\bibinfo  {journal} {Phys. Rev. Lett.}\ }\textbf {\bibinfo {volume} {97}},\
	\bibinfo {pages} {026603} (\bibinfo {year} {2006})}\BibitemShut {NoStop}%
\bibitem [{\citenamefont {Kaplan}\ \emph {et~al.}(2023)\citenamefont {Kaplan},
	\citenamefont {Holder},\ and\ \citenamefont {Yan}}]{Kaplan2023}%
\BibitemOpen
\bibfield  {author} {\bibinfo {author} {\bibfnamefont {D.}~\bibnamefont
		{Kaplan}}, \bibinfo {author} {\bibfnamefont {T.}~\bibnamefont {Holder}},\
	and\ \bibinfo {author} {\bibfnamefont {B.}~\bibnamefont {Yan}},\ }\bibfield
{title} {\bibinfo {title} {Unifying semiclassics and quantum perturbation
		theory at nonlinear order},\ }\href
{https://doi.org/10.21468/scipostphys.14.4.082} {\bibfield  {journal}
	{\bibinfo  {journal} {SciPost Phys.}\ }\textbf {\bibinfo {volume} {14}},\
	\bibinfo {pages} {082} (\bibinfo {year} {2023})}\BibitemShut {NoStop}%
\bibitem [{\citenamefont {Lai}\ \emph {et~al.}(2021)\citenamefont {Lai},
	\citenamefont {Liu}, \citenamefont {Zhang}, \citenamefont {Zhao},
	\citenamefont {Feng}, \citenamefont {Wang}, \citenamefont {Tang},
	\citenamefont {Liu}, \citenamefont {Novoselov}, \citenamefont {Yang},\ and\
	\citenamefont {Gao}}]{Lai2021}%
\BibitemOpen
\bibfield  {author} {\bibinfo {author} {\bibfnamefont {S.}~\bibnamefont
		{Lai}}, \bibinfo {author} {\bibfnamefont {H.}~\bibnamefont {Liu}}, \bibinfo
	{author} {\bibfnamefont {Z.}~\bibnamefont {Zhang}}, \bibinfo {author}
	{\bibfnamefont {J.}~\bibnamefont {Zhao}}, \bibinfo {author} {\bibfnamefont
		{X.}~\bibnamefont {Feng}}, \bibinfo {author} {\bibfnamefont {N.}~\bibnamefont
		{Wang}}, \bibinfo {author} {\bibfnamefont {C.}~\bibnamefont {Tang}}, \bibinfo
	{author} {\bibfnamefont {Y.}~\bibnamefont {Liu}}, \bibinfo {author}
	{\bibfnamefont {K.~S.}\ \bibnamefont {Novoselov}}, \bibinfo {author}
	{\bibfnamefont {S.~A.}\ \bibnamefont {Yang}},\ and\ \bibinfo {author}
	{\bibfnamefont {W.-b.}\ \bibnamefont {Gao}},\ }\bibfield  {title} {\bibinfo
	{title} {Third-order nonlinear Hall effect induced by the Berry-connection
		polarizability tensor},\ }\href {https://doi.org/10.1038/s41565-021-00917-0}
{\bibfield  {journal} {\bibinfo  {journal} {Nat. Nanotechnol.}\ }\textbf
	{\bibinfo {volume} {16}},\ \bibinfo {pages} {869} (\bibinfo {year}
	{2021})}\BibitemShut {NoStop}%
\bibitem [{\citenamefont {Tang}\ \emph {et~al.}(2016)\citenamefont {Tang},
	\citenamefont {Zhou}, \citenamefont {Xu},\ and\ \citenamefont
	{Zhang}}]{Tang2016}%
\BibitemOpen
\bibfield  {author} {\bibinfo {author} {\bibfnamefont {P.}~\bibnamefont
		{Tang}}, \bibinfo {author} {\bibfnamefont {Q.}~\bibnamefont {Zhou}}, \bibinfo
	{author} {\bibfnamefont {G.}~\bibnamefont {Xu}},\ and\ \bibinfo {author}
	{\bibfnamefont {S.-C.}\ \bibnamefont {Zhang}},\ }\bibfield  {title} {\bibinfo
	{title} {Dirac fermions in an antiferromagnetic semimetal},\ }\href
{https://doi.org/10.1038/nphys3839} {\bibfield  {journal} {\bibinfo
		{journal} {Nat. Phys.}\ }\textbf {\bibinfo {volume} {12}},\ \bibinfo {pages}
	{1100} (\bibinfo {year} {2016})}\BibitemShut {NoStop}%
\bibitem [{\citenamefont {Sinitsyn}\ \emph {et~al.}(2005)\citenamefont
	{Sinitsyn}, \citenamefont {Niu}, \citenamefont {Sinova},\ and\ \citenamefont
	{Nomura}}]{Sinitsyn2005}%
\BibitemOpen
\bibfield  {author} {\bibinfo {author} {\bibfnamefont {N.~A.}\ \bibnamefont
		{Sinitsyn}}, \bibinfo {author} {\bibfnamefont {Q.}~\bibnamefont {Niu}},
	\bibinfo {author} {\bibfnamefont {J.}~\bibnamefont {Sinova}},\ and\ \bibinfo
	{author} {\bibfnamefont {K.}~\bibnamefont {Nomura}},\ }\bibfield  {title}
{\bibinfo {title} {Disorder effects in the anomalous Hall effect induced by
		Berry curvature},\ }\href {https://doi.org/10.1103/physrevb.72.045346}
{\bibfield  {journal} {\bibinfo  {journal} {Phys. Rev. B}\ }\textbf {\bibinfo
		{volume} {72}},\ \bibinfo {pages} {045346} (\bibinfo {year}
	{2005})}\BibitemShut {NoStop}%
\bibitem [{\citenamefont {Du}\ \emph {et~al.}(2019)\citenamefont {Du},
	\citenamefont {Wang}, \citenamefont {Li}, \citenamefont {Lu},\ and\
	\citenamefont {Xie}}]{Du2019}%
\BibitemOpen
\bibfield  {author} {\bibinfo {author} {\bibfnamefont {Z.~Z.}\ \bibnamefont
		{Du}}, \bibinfo {author} {\bibfnamefont {C.~M.}\ \bibnamefont {Wang}},
	\bibinfo {author} {\bibfnamefont {S.}~\bibnamefont {Li}}, \bibinfo {author}
	{\bibfnamefont {H.-Z.}\ \bibnamefont {Lu}},\ and\ \bibinfo {author}
	{\bibfnamefont {X.~C.}\ \bibnamefont {Xie}},\ }\bibfield  {title} {\bibinfo
	{title} {Disorder-induced nonlinear Hall effect with time-reversal
		symmetry},\ }\href {https://doi.org/10.1038/s41467-019-10941-3} {\bibfield
	{journal} {\bibinfo  {journal} {Nat. Commun.}\ }\textbf {\bibinfo {volume}
		{10}},\ \bibinfo {pages} {3047} (\bibinfo {year} {2019})}\BibitemShut
{NoStop}%
\bibitem [{\citenamefont {Atencia}\ \emph {et~al.}(2023)\citenamefont
	{Atencia}, \citenamefont {Xiao},\ and\ \citenamefont {Culcer}}]{Atencia2023}%
\BibitemOpen
\bibfield  {author} {\bibinfo {author} {\bibfnamefont {R.~B.}\ \bibnamefont
		{Atencia}}, \bibinfo {author} {\bibfnamefont {D.}~\bibnamefont {Xiao}},\ and\
	\bibinfo {author} {\bibfnamefont {D.}~\bibnamefont {Culcer}},\ }\bibfield
{title} {\bibinfo {title} {Disorder in the nonlinear anomalous Hall effect of
		PT-symmetric dirac fermions},\ }\href
{https://doi.org/10.1103/physrevb.108.l201115} {\bibfield  {journal}
	{\bibinfo  {journal} {Phys. Rev. B}\ }\textbf {\bibinfo {volume} {108}},\
	\bibinfo {pages} {l201115} (\bibinfo {year} {2023})}\BibitemShut {NoStop}%
\bibitem [{\citenamefont {Arai}\ \emph {et~al.}(2024)\citenamefont {Arai},
	\citenamefont {Hayashi}, \citenamefont {Takeda}, \citenamefont {Tou},
	\citenamefont {Sugawara},\ and\ \citenamefont {Kotegawa}}]{Arai2024}%
\BibitemOpen
\bibfield  {author} {\bibinfo {author} {\bibfnamefont {Y.}~\bibnamefont
		{Arai}}, \bibinfo {author} {\bibfnamefont {J.}~\bibnamefont {Hayashi}},
	\bibinfo {author} {\bibfnamefont {K.}~\bibnamefont {Takeda}}, \bibinfo
	{author} {\bibfnamefont {H.}~\bibnamefont {Tou}}, \bibinfo {author}
	{\bibfnamefont {H.}~\bibnamefont {Sugawara}},\ and\ \bibinfo {author}
	{\bibfnamefont {H.}~\bibnamefont {Kotegawa}},\ }\bibfield  {title} {\bibinfo
	{title} {Intrinsic anomalous Hall effect arising from antiferromagnetism as
		revealed by high-quality NbMnP},\ }\href
{https://doi.org/10.7566/jpsj.93.063702} {\bibfield  {journal} {\bibinfo
		{journal} {J. Phys. Soc. Jpn.}\ }\textbf {\bibinfo {volume} {93}},\ \bibinfo
	{pages} {063702} (\bibinfo {year} {2024})}\BibitemShut {NoStop}%
\bibitem [{\citenamefont {Ding}\ \emph {et~al.}(2019)\citenamefont {Ding},
	\citenamefont {Koo}, \citenamefont {Xu}, \citenamefont {Li}, \citenamefont
	{Lu}, \citenamefont {Zhao}, \citenamefont {Wang}, \citenamefont {Yin},
	\citenamefont {Lei}, \citenamefont {Yan}, \citenamefont {Zhu},\ and\
	\citenamefont {Behnia}}]{Ding2019}%
\BibitemOpen
\bibfield  {author} {\bibinfo {author} {\bibfnamefont {L.}~\bibnamefont
		{Ding}}, \bibinfo {author} {\bibfnamefont {J.}~\bibnamefont {Koo}}, \bibinfo
	{author} {\bibfnamefont {L.}~\bibnamefont {Xu}}, \bibinfo {author}
	{\bibfnamefont {X.}~\bibnamefont {Li}}, \bibinfo {author} {\bibfnamefont
		{X.}~\bibnamefont {Lu}}, \bibinfo {author} {\bibfnamefont {L.}~\bibnamefont
		{Zhao}}, \bibinfo {author} {\bibfnamefont {Q.}~\bibnamefont {Wang}}, \bibinfo
	{author} {\bibfnamefont {Q.}~\bibnamefont {Yin}}, \bibinfo {author}
	{\bibfnamefont {H.}~\bibnamefont {Lei}}, \bibinfo {author} {\bibfnamefont
		{B.}~\bibnamefont {Yan}}, \bibinfo {author} {\bibfnamefont {Z.}~\bibnamefont
		{Zhu}},\ and\ \bibinfo {author} {\bibfnamefont {K.}~\bibnamefont {Behnia}},\
}\bibfield  {title} {\bibinfo {title} {Intrinsic anomalous Nernst effect
		amplified by disorder in a half-metallic semimetal},\ }\href
{https://doi.org/10.1103/physrevx.9.041061} {\bibfield  {journal} {\bibinfo
		{journal} {Phys. Rev. X}\ }\textbf {\bibinfo {volume} {9}},\ \bibinfo {pages}
	{041061} (\bibinfo {year} {2019})}\BibitemShut {NoStop}%
\bibitem [{\citenamefont {Gao}\ \emph {et~al.}(2023)\citenamefont {Gao},
	\citenamefont {Liu}, \citenamefont {Qiu}, \citenamefont {Ghosh},
	\citenamefont {V.~Trevisan}, \citenamefont {Onishi}, \citenamefont {Hu},
	\citenamefont {Qian}, \citenamefont {Tien}, \citenamefont {Chen},
	\citenamefont {Huang}, \citenamefont {Bérubé}, \citenamefont {Li},
	\citenamefont {Tzschaschel}, \citenamefont {Dinh}, \citenamefont {Sun},
	\citenamefont {Ho}, \citenamefont {Lien}, \citenamefont {Singh},
	\citenamefont {Watanabe}, \citenamefont {Taniguchi}, \citenamefont {Bell},
	\citenamefont {Lin}, \citenamefont {Chang}, \citenamefont {Du}, \citenamefont
	{Bansil}, \citenamefont {Fu}, \citenamefont {Ni}, \citenamefont {Orth},
	\citenamefont {Ma},\ and\ \citenamefont {Xu}}]{Gao2023}%
\BibitemOpen
\bibfield  {author} {\bibinfo {author} {\bibfnamefont {A.}~\bibnamefont
		{Gao}}, \bibinfo {author} {\bibfnamefont {Y.-F.}\ \bibnamefont {Liu}},
	\bibinfo {author} {\bibfnamefont {J.-X.}\ \bibnamefont {Qiu}}, \bibinfo
	{author} {\bibfnamefont {B.}~\bibnamefont {Ghosh}}, \bibinfo {author}
	{\bibfnamefont {T.}~\bibnamefont {V.~Trevisan}}, \bibinfo {author}
	{\bibfnamefont {Y.}~\bibnamefont {Onishi}}, \bibinfo {author} {\bibfnamefont
		{C.}~\bibnamefont {Hu}}, \bibinfo {author} {\bibfnamefont {T.}~\bibnamefont
		{Qian}}, \bibinfo {author} {\bibfnamefont {H.-J.}\ \bibnamefont {Tien}},
	\bibinfo {author} {\bibfnamefont {S.-W.}\ \bibnamefont {Chen}}, \bibinfo
	{author} {\bibfnamefont {M.}~\bibnamefont {Huang}}, \bibinfo {author}
	{\bibfnamefont {D.}~\bibnamefont {Bérubé}}, \bibinfo {author}
	{\bibfnamefont {H.}~\bibnamefont {Li}}, \bibinfo {author} {\bibfnamefont
		{C.}~\bibnamefont {Tzschaschel}}, \bibinfo {author} {\bibfnamefont
		{T.}~\bibnamefont {Dinh}}, \bibinfo {author} {\bibfnamefont {Z.}~\bibnamefont
		{Sun}}, \bibinfo {author} {\bibfnamefont {S.-C.}\ \bibnamefont {Ho}},
	\bibinfo {author} {\bibfnamefont {S.-W.}\ \bibnamefont {Lien}}, \bibinfo
	{author} {\bibfnamefont {B.}~\bibnamefont {Singh}}, \bibinfo {author}
	{\bibfnamefont {K.}~\bibnamefont {Watanabe}}, \bibinfo {author}
	{\bibfnamefont {T.}~\bibnamefont {Taniguchi}}, \bibinfo {author}
	{\bibfnamefont {D.~C.}\ \bibnamefont {Bell}}, \bibinfo {author}
	{\bibfnamefont {H.}~\bibnamefont {Lin}}, \bibinfo {author} {\bibfnamefont
		{T.-R.}\ \bibnamefont {Chang}}, \bibinfo {author} {\bibfnamefont {C.~R.}\
		\bibnamefont {Du}}, \bibinfo {author} {\bibfnamefont {A.}~\bibnamefont
		{Bansil}}, \bibinfo {author} {\bibfnamefont {L.}~\bibnamefont {Fu}}, \bibinfo
	{author} {\bibfnamefont {N.}~\bibnamefont {Ni}}, \bibinfo {author}
	{\bibfnamefont {P.~P.}\ \bibnamefont {Orth}}, \bibinfo {author}
	{\bibfnamefont {Q.}~\bibnamefont {Ma}},\ and\ \bibinfo {author}
	{\bibfnamefont {S.-Y.}\ \bibnamefont {Xu}},\ }\bibfield  {title} {\bibinfo
	{title} {Quantum metric nonlinear Hall effect in a topological
		antiferromagnetic heterostructure},\ }\href
{https://doi.org/10.1126/science.adf1506} {\bibfield  {journal} {\bibinfo
		{journal} {Science}\ }\textbf {\bibinfo {volume} {381}},\ \bibinfo {pages}
	{181} (\bibinfo {year} {2023})}\BibitemShut {NoStop}%
\bibitem [{\citenamefont {Deng}\ \emph {et~al.}(2020)\citenamefont {Deng},
	\citenamefont {Yu}, \citenamefont {Shi}, \citenamefont {Guo}, \citenamefont
	{Xu}, \citenamefont {Wang}, \citenamefont {Chen},\ and\ \citenamefont
	{Zhang}}]{Deng2020}%
\BibitemOpen
\bibfield  {author} {\bibinfo {author} {\bibfnamefont {Y.}~\bibnamefont
		{Deng}}, \bibinfo {author} {\bibfnamefont {Y.}~\bibnamefont {Yu}}, \bibinfo
	{author} {\bibfnamefont {M.~Z.}\ \bibnamefont {Shi}}, \bibinfo {author}
	{\bibfnamefont {Z.}~\bibnamefont {Guo}}, \bibinfo {author} {\bibfnamefont
		{Z.}~\bibnamefont {Xu}}, \bibinfo {author} {\bibfnamefont {J.}~\bibnamefont
		{Wang}}, \bibinfo {author} {\bibfnamefont {X.~H.}\ \bibnamefont {Chen}},\
	and\ \bibinfo {author} {\bibfnamefont {Y.}~\bibnamefont {Zhang}},\ }\bibfield
{title} {\bibinfo {title} {Quantum anomalous Hall effect in intrinsic
		magnetic topological insulator MnBi$_2$Te$_4$},\ }\href
{https://doi.org/10.1126/science.aax8156} {\bibfield  {journal} {\bibinfo
		{journal} {Science}\ }\textbf {\bibinfo {volume} {367}},\ \bibinfo {pages}
	{895} (\bibinfo {year} {2020})}\BibitemShut {NoStop}%
\bibitem [{\citenamefont {Li}\ \emph {et~al.}(2024)\citenamefont {Li},
	\citenamefont {Zhang}, \citenamefont {Zhou}, \citenamefont {Ma},
	\citenamefont {Lei}, \citenamefont {Jin}, \citenamefont {He}, \citenamefont
	{Li}, \citenamefont {Law},\ and\ \citenamefont {Wang}}]{Li2024a}%
\BibitemOpen
\bibfield  {author} {\bibinfo {author} {\bibfnamefont {H.}~\bibnamefont
		{Li}}, \bibinfo {author} {\bibfnamefont {C.}~\bibnamefont {Zhang}}, \bibinfo
	{author} {\bibfnamefont {C.}~\bibnamefont {Zhou}}, \bibinfo {author}
	{\bibfnamefont {C.}~\bibnamefont {Ma}}, \bibinfo {author} {\bibfnamefont
		{X.}~\bibnamefont {Lei}}, \bibinfo {author} {\bibfnamefont {Z.}~\bibnamefont
		{Jin}}, \bibinfo {author} {\bibfnamefont {H.}~\bibnamefont {He}}, \bibinfo
	{author} {\bibfnamefont {B.}~\bibnamefont {Li}}, \bibinfo {author}
	{\bibfnamefont {K.~T.}\ \bibnamefont {Law}},\ and\ \bibinfo {author}
	{\bibfnamefont {J.}~\bibnamefont {Wang}},\ }\bibfield  {title} {\bibinfo
	{title} {Quantum geometry quadrupole-induced third-order nonlinear transport
		in antiferromagnetic topological insulator MnBi$_2$Te$_4$},\ }\href
{https://doi.org/10.1038/s41467-024-52206-8} {\bibfield  {journal} {\bibinfo
		{journal} {Nat. Commun.}\ }\textbf {\bibinfo {volume} {15}},\ \bibinfo
	{pages} {7779} (\bibinfo {year} {2024})}\BibitemShut {NoStop}%
\bibitem [{\citenamefont {He}(2020)}]{He2020}%
\BibitemOpen
\bibfield  {author} {\bibinfo {author} {\bibfnamefont {K.}~\bibnamefont
		{He}},\ }\bibfield  {title} {\bibinfo {title} {MnBi$_2$Te$_4$-family intrinsic
		magnetic topological materials},\ }\href
{https://doi.org/10.1038/s41535-020-00291-5} {\bibfield  {journal} {\bibinfo
		{journal} {npj Quantum Materials}\ }\textbf {\bibinfo {volume} {5}},\
	\bibinfo {pages} {90} (\bibinfo {year} {2020})}\BibitemShut {NoStop}%
\bibitem [{\citenamefont {Lin}\ \emph {et~al.}(2022)\citenamefont {Lin},
	\citenamefont {Feng}, \citenamefont {Wang}, \citenamefont {Zhu},
	\citenamefont {Lian}, \citenamefont {Zhang}, \citenamefont {Li},
	\citenamefont {Wu}, \citenamefont {Liu}, \citenamefont {Wang}, \citenamefont
	{Zhang}, \citenamefont {Wang}, \citenamefont {Chen}, \citenamefont {Zhou},\
	and\ \citenamefont {Shen}}]{Lin2022}%
\BibitemOpen
\bibfield  {author} {\bibinfo {author} {\bibfnamefont {W.}~\bibnamefont
		{Lin}}, \bibinfo {author} {\bibfnamefont {Y.}~\bibnamefont {Feng}}, \bibinfo
	{author} {\bibfnamefont {Y.}~\bibnamefont {Wang}}, \bibinfo {author}
	{\bibfnamefont {J.}~\bibnamefont {Zhu}}, \bibinfo {author} {\bibfnamefont
		{Z.}~\bibnamefont {Lian}}, \bibinfo {author} {\bibfnamefont {H.}~\bibnamefont
		{Zhang}}, \bibinfo {author} {\bibfnamefont {H.}~\bibnamefont {Li}}, \bibinfo
	{author} {\bibfnamefont {Y.}~\bibnamefont {Wu}}, \bibinfo {author}
	{\bibfnamefont {C.}~\bibnamefont {Liu}}, \bibinfo {author} {\bibfnamefont
		{Y.}~\bibnamefont {Wang}}, \bibinfo {author} {\bibfnamefont {J.}~\bibnamefont
		{Zhang}}, \bibinfo {author} {\bibfnamefont {Y.}~\bibnamefont {Wang}},
	\bibinfo {author} {\bibfnamefont {C.-Z.}\ \bibnamefont {Chen}}, \bibinfo
	{author} {\bibfnamefont {X.}~\bibnamefont {Zhou}},\ and\ \bibinfo {author}
	{\bibfnamefont {J.}~\bibnamefont {Shen}},\ }\bibfield  {title} {\bibinfo
	{title} {Direct visualization of edge state in even-layer MnBi$_2$Te$_4$ at zero
		magnetic field},\ }\href {https://doi.org/10.1038/s41467-022-35482-0}
{\bibfield  {journal} {\bibinfo  {journal} {Nat. Commun.}\ }\textbf {\bibinfo
		{volume} {13}},\ \bibinfo {pages} {7714} (\bibinfo {year}
	{2022})}\BibitemShut {NoStop}%
\bibitem [{\citenamefont {Li}\ \emph {et~al.}(2023)\citenamefont {Li},
	\citenamefont {Liu}, \citenamefont {Liu}, \citenamefont {Wang}, \citenamefont
	{Lu},\ and\ \citenamefont {Xie}}]{Li2023}%
\BibitemOpen
\bibfield  {author} {\bibinfo {author} {\bibfnamefont {S.}~\bibnamefont
		{Li}}, \bibinfo {author} {\bibfnamefont {T.}~\bibnamefont {Liu}}, \bibinfo
	{author} {\bibfnamefont {C.}~\bibnamefont {Liu}}, \bibinfo {author}
	{\bibfnamefont {Y.}~\bibnamefont {Wang}}, \bibinfo {author} {\bibfnamefont
		{H.-Z.}\ \bibnamefont {Lu}},\ and\ \bibinfo {author} {\bibfnamefont {X.~C.}\
		\bibnamefont {Xie}},\ }\bibfield  {title} {\bibinfo {title} {Progress on the
		antiferromagnetic topological insulator MnBi$_2$Te$_4$},\ }\href
{https://doi.org/10.1093/nsr/nwac296} {\bibfield  {journal} {\bibinfo
		{journal} {Natl. Sci. Rev.}\ }\textbf {\bibinfo {volume} {11}},\ \bibinfo
	{pages} {nwac296} (\bibinfo {year} {2023})}\BibitemShut {NoStop}%
	\end{thebibliography}
	
\end{document}